\documentclass[journal=jpcbfk,manuscript=article]{achemso}
\usepackage{textcomp}
\usepackage{longtable}
\usepackage{multirow}
\usepackage{amsmath}
\usepackage{subfigure}
\usepackage{bm}
\usepackage[usenames,dvipsnames]{color}
\usepackage{soul, xcolor}
\usepackage{threeparttable}
\usepackage[normalem]{ulem}

\usepackage{xr}
\SectionNumbersOn

\author{Silvan K\"aser} \affiliation[University of Basel]{Department
  of Chemistry, University of Basel, Klingelbergstrasse 80 , CH-4056
  Basel, Switzerland.}  \author{Oliver T. Unke} \affiliation[TU
  Berlin]{Department of Chemistry, University of Basel,
  Klingelbergstrasse 80, CH-4056 Basel, Switzerland\\ Present Address:
  Machine Learning Group, TU Berlin, Marchstr. 23, 10587 Berlin,
  Germany} \author{Markus Meuwly} \affiliation[University of
  Basel]{Department of Chemistry, University of Basel,
  Klingelbergstrasse 80 , CH-4056 Basel, Switzerland.}
\email{m.meuwly@unibas.ch}

\title {Reactive Dynamics and Spectroscopy of Hydrogen Transfer from
  Neural Network-Based Reactive Potential Energy Surfaces}

\begin{document}

\date{\today}

\begin{abstract}
The ``in silico'' exploration of chemical, physical and biological
systems requires accurate and efficient energy functions to follow
their nuclear dynamics at a molecular and atomistic level. Recently,
machine learning tools gained a lot of attention in the field of
molecular sciences and simulations and are increasingly used to
investigate the dynamics of such systems. Among the various
approaches, artificial neural networks (NNs) are one promising tool to
learn a representation of potential energy surfaces. This is done by
formulating the problem as a mapping from a set of atomic positions
$\mathbf{x}$ and nuclear charges $Z_i$ to a potential energy
$V(\mathbf{x})$. Here, a fully-dimensional, reactive neural network
representation for malonaldehyde (MA), acetoacetaldehyde (AAA) and
acetylacetone (AcAc) is learned. It is used to run finite-temperature
molecular dynamics simulations, and to determine the infrared spectra
and the hydrogen transfer rates for the three molecules. The
finite-temperature infrared spectrum for MA based on the NN learned on
MP2 reference data provides a realistic representation of the
low-frequency modes and the H-transfer band whereas the CH vibrations
are somewhat too high in frequency. For AAA it is demonstrated that
the IR spectroscopy is sensitive to the position of the transferring
hydrogen at either the OCH- or OCCH$_3$ end of the molecule. For the
hydrogen transfer rates it is demonstrated that the O--O vibration (at
$\sim 250$ cm$^{-1}$) is a gating mode and largely determines the rate
at which the hydrogen is transferred between the donor and
acceptor. Finally, possibilities to further improve such NN-based
potential energy surfaces are explored. They include the
transferability of an NN-learned energy function across chemical
species (here methylation) and transfer learning from a lower level of
reference data (MP2) to a higher level of theory (pair natural
orbital-LCCSD(T)).
\end{abstract}

\section{Introduction}
Rapid progress of computer technology has given science new
opportunities. With the possibility to carry out simulations in the
broadest sense, the conventional approach to research consisting of
experimental and theoretical/mathematical methods has been
considerably extended. This is particularly true for the molecular
sciences for which realistic, atomically-resolved simulations have
become possible in the past two decades. Such ``in silico'' approaches
now constitute an integral part of the toolbox of molecular scientists
to characterize, formulate and test hypotheses and make predictions
about complex systems.\cite{qiang.jcp:2016,MM.rev.sd:2017}\\

\noindent
A necessary requirement for carrying out such atomistic simulations --
both, molecular dynamics (MD) and Monte Carlo (MC) -- is the
availability of a means to determine the total energy of the system
for given positions $\mathbf{x}$ of all the atoms. For MD simulations,
the forces are required as well. The most rigorous approach would be
to recompute the total energy for every new configuration using
electronic structure methods, i.e. solving the electronic
Schr\"odinger equation at the highest level of theory and with the
largest basis set that is computationally affordable. However, this is
impractical even for small systems due to a number of reasons, in
particular if a statistically significant number of trajectories is
required. First, the computing time for high-level quantum methods is
appreciable. Second, technical aspects such as accounting for basis
set superposition errors or including multi reference effects for
highly distorted geometries are difficult. Finally, for chemically
demanding systems, e.g. those containing metal ions, it may even be
difficult to converge the Hartree-Fock wavefunction to the desired
state or converging it at all for arbitrary geometries. In such cases,
human intervention is required which is not desirable.\\

\noindent
As an alternative, the total energy can be computed for a number of
conformations {\it a priori} on a grid. Then, the potential energy
surface (PES) needs to be represented in a way that can be evaluated
for arbitrary geometries. This can be done by either resorting to a
fit of a parametrized function or by representing the PES, e.g. using
reproducing kernel Hilbert space
theory.\cite{hollebeek.annrevphychem.1999.rkhs,MM.rkhs:2017}
Alternatively, machine learning (ML) methods have recently emerged as
a possibility to represent the energetics of molecules and their
intermolecular interactions.\cite{behler2014representing} Here, the
energies are obtained from learning a representation of the potential
energy surface (PES) of a system, which connects a set of nuclear
charges and atomic positions to the energy.\cite{unke2019physnet} A
suitable tool for ``learning'' molecular energies are so-called
artificial neural networks (ANNs, henceforth NNs), which were shown to
be general function approximators.\cite{hornik1989multilayer}\\

\begin{figure}[htbp]
\centering
\includegraphics[width=0.7\textwidth]{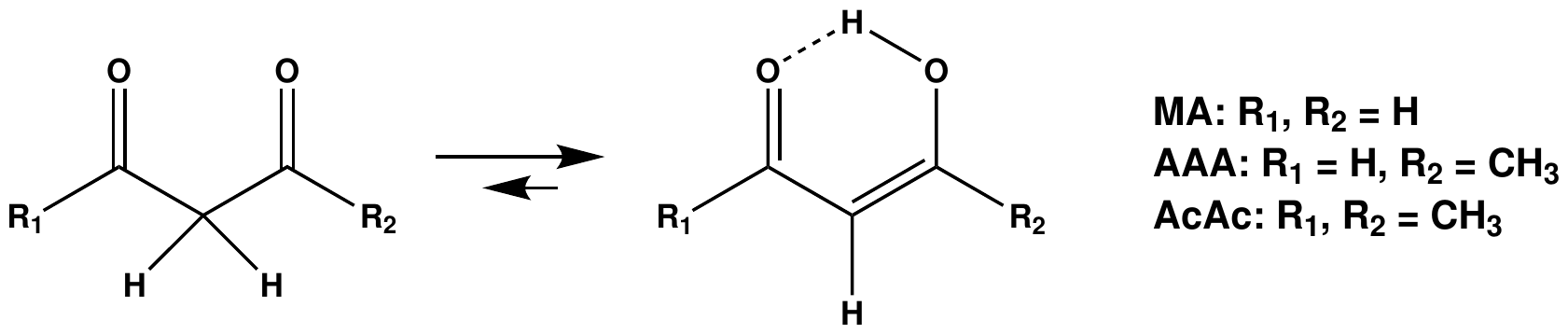}
\caption{Tautomerization of a $\beta$-diketone}
\label{fig:structures}
\end{figure}

\noindent
In the present work, an NN based on the PhysNet \cite{unke2019physnet}
architecture is used to generate and explore fully-dimensional,
reactive PESs for three $\beta$-diketones, see Figure
\ref{fig:structures}. The structures all have an enol- and a
keto-form, of which the enol form is more stable in the gas
phase.\cite{yamabe2004reaction} Malonaldehyde (MA,
\textit{propandial}) is a well-studied system (see below), both
experimentally and by computation for hydrogen transfer (HT) which is
one of the ubiquitous reaction mechanisms in
chemistry.\cite{caldin2013proton} The other two systems,
acetoacetaldehyde (AAA, \textit{3-oxobutanal}) and acetylacetone
(AcAc, \textit{pentan-2,4-dion}) examine the influence of
methyl-substituents on the intramolecular HT for both, symmetric
(AcAc) and asymmetric (AAA) substitution.\\

\noindent
For MA, the infrared (IR)
spectrum\cite{maexp83,maexp89,maexp92,maexp04} and the ground state
tunneling splitting\cite{mawilson81,mafirth91,maexp10} for hydrogen
transfer have been determined experimentally. Calculations at
different levels of theory have also been carried out to assign these
spectra and to reproduce the splitting. In general, this requires
fully dimensional dynamics
simulations~\cite{macal07,macal09,ma11,MM10ma,macal11,MM.ma:2014}
performed on high level PESs.~\cite{macal08}\\

\noindent
AcAc is structurally related to MA (see Figure \ref{fig:structures})
through substitution of the symmetrical H-atoms by methyl
groups. Given that the methyl torsion can couple to the O--O stretch
and hence to the hydrogen motion along the H--bond, it constitutes a
more challenging problem than HT in MA. In AAA, only one of the
hydrogen atoms is replaced by methyl.\\

\noindent
Contrary to MA, less information about activation barriers, structures
and possible tunneling splittings in AcAc is available. Even the
question whether its ground state assumes an asymmetric ($C_{s}$) or a
symmetric ($C_{2v}$) structure is still
debated.\cite{JOH:JCP02,LOW:JACS71,AND:JMS72,SRI:JACS04,IIJ:JMS87,CAM:JACS06}
The ground state structure from neutron crystallography predicts a
$C_{s}$ symmetry\cite{JOH:JCP02} whereas electron diffraction
experiments suggest either a $C_{s}$\cite{LOW:JACS71,AND:JMS72} or a
$C_{2v}$ structure.\cite{IIJ:JMS87} The most recent study performed
with ultrafast electron diffraction concluded that the lowest energy
form of AcAc has $C_s$ symmetry.\cite{SRI:JACS04} In general,
electronic structure calculations find an asymmetric minimum energy
structure with $C_s$ symmetry for AcAc on the PESs excluding
zero-point
corrections,\cite{BAU:CPL97,SLI:JSC02,MAT:CP04,MAT:JPC05,CAM:JMS05,DAN:JPC94}
while correcting for zero point vibrational energy leads to a slight
preference of a $C_{2v}$ structure.\cite{DAN:JPC94} \\

\noindent
Conversely, high-resolution rotational spectra of AcAc and its singly
substituted $^{13}$C-iso\-to\-po\-logues in the frequency ranges 2--26.5 GHz
($\sim$0.1--1 cm$^{-1}$) and 60--80 GHz \mbox{($\sim$2--3 cm$^{-1}$)}
lead to a structure with $C_{2v}$ symmetry~\cite{CAM:JACS06} possibly
due to zero point vibration of the proton in systems with strong
hydrogen bonds.~\cite{asmis07,yang08jcp,yang08zpc} The double well
potential in AcAc due to HT from one oxygen to another is responsible
for the tunneling splitting.\cite{HIN:JCP97,MAV:JPC01}\\

\noindent
Several IR spectroscopic studies of AcAc are available in the vapor
phase.\cite{OGO:JCP66,FUN:HB59,TAY:SA79,TAY:SA00} The OH--stretching
transition is very broad, however, it is typically assigned in the
region from 2750 cm$^{-1}$ to 2800 cm$^{-1}$.\cite{OGO:JCP66,TAY:SA00}
The only near infrared (NIR) investigation of AcAc was performed in
the liquid phase in 1929 as part of a study regarding carbonyl
overtones.\cite{ELL:JACS29} Other than a tentative assignment of a
carbonyl overtone at 1.91 \textmu m, no further transitions for AcAc
were assigned. More recently, the vapor phase IR and NIR spectra of
AcAc were reported.\cite{MM.acac:2015} The experimental spectra were
interpreted by using atomistic simulations combined with reactive
force fields based on molecular mechanics with proton transfer
(MMPT).\cite{MM.mmpt:2008} To the best of our knowledge no
spectroscopic data for AAA is available in the open literature.\\

\noindent
In the present work fully-dimensional, reactive PESs for MA, AAA, and
AcAc are constructed by training a single NN-representation using
electronic structure data. The PESs are used in finite-temperature MD
simulations to determine the IR spectra of the three compounds and the
(classical) rates for HT. Machine learning of the PESs is also carried
out in two further, complementary ways. First, the generalizability of
such an NN-learned PES is tested by learning the AcAc PES based on
reference data only from MA, AAA, and the underlying
``amons''\cite{huang2017amons}. Second, transfer learning
\cite{pan2009survey} (TL) of an NN trained on lower-level {\it ab
  initio} data to a higher level of theory is attempted.\\

\noindent
The work is structured as follows. First, the methods are
discussed. Then, the results from finite-temperature MD simulations
for the three systems are presented and discussed in the context of
previous experimental and computational work. This is followed by
generalizations of the NN-learned PES and the results for TL. Finally,
conclusions are drawn and an outlook is given.\\

\section{Methods}

\subsection{Neural Network PESs Based on PhysNet}
All PESs used in this work are constructed by fitting the parameters
of the PhysNet\cite{unke2019physnet} NN architecture to extensive
reference \textit{ab initio} data. The NN architecture, the data set
generation process and the training (fitting) procedure are described
below.\\

\noindent
{\it Neural network architecture:} PhysNet\cite{unke2019physnet} is a
high-dimensional NN\cite{behler2007generalized} of the
``message-passing''\cite{gilmer2017message} type. It predicts atomic
energy contributions and partial charges from feature vectors encoding
information about the local chemical environment of each
atom.\cite{unke2018reactive} The feature vectors are constructed
iteratively from the nuclear charges $Z_i$ and Cartesian coordinates
$\mathbf{r}_i$ of all atoms $i$ by passing ``messages'' between atoms
within a cut-off distance $r_{\rm cut}$ (here $r_{\rm cut} =
10$~\r{A}). Long-range electrostatics and dispersion interactions are
included explicitly. The total potential energy $E$ of a system is
represented as
\begin{equation}
E =\sum_{i=1}^N E_i + k_e\sum_{i=1}^N\sum_{j>i}^N \frac{q_i
  q_j}{r_{ij}}+E_{\rm D3}
\label{eq:physnet_total_energy}
\end{equation}
where $E_i$ and $q_i$ are the predicted atomic energy contributions
and partial charges of atom $i$, $r_{ij}$ is the distance between
atoms $i$ and $j$, $k_e$ is Coulomb's constant and $E_{\rm D3}$ is the
D3 dispersion correction\cite{grimme2010consistent}. To avoid
numerical instabilities, the Coulomb term is damped for small
distances $r_{ij}$ (not shown in Eq.~\ref{eq:physnet_total_energy} for
simplicity, see Ref.~\citenum{unke2019physnet} for details) and charge
neutrality of the molecule is explicitly enforced, see
Ref.~\citenum{unke2019physnet}. The functional form for the atom
embedding is invariant with respect to translation, rotation and
permutation of atoms sharing the same nuclear charge
$Z$.\cite{unke2019physnet}\\

\noindent
The forces $\mathbf{F}_i$ acting on each atom $i$, required for MD
studies, can be obtained analytically by reverse mode automatic
differentiation\cite{baydin2018automatic}. For computing the infrared
(IR) spectra based on the MD simulations, the total molecular dipole
moment is computed from the  (fluctuating) partial
charges $q_i$ (Eq.~\ref{eq:physnet_dipole_moment}).
\begin{equation}
\boldsymbol{\mu} = \sum_{i=1}^N q_i\mathbf{r}_i
\label{eq:physnet_dipole_moment}
\end{equation}

\noindent
During the NN training the parameters are fitted to reference
\textit{ab initio} energies, forces and dipole moments using the
procedure described in Ref.~\citenum{unke2019physnet}. The parameters
for the D3 dispersion correction were initialized to standard values
for the Hartree-Fock method \cite{grimme2011effect} and then refined
during training. Minimization was carried out using ``adaptive moment
estimation'' (Adam)\cite{adam:2014} which combines momentum and root
mean squared propagation.\\

\noindent
{\it Data set generation:} The quality of the reference data set is
crucial for generating an accurate and robust PES. Here, the general
procedure followed Ref.~\citenum{unke2019physnet} again. The data set
for the present work is based on calculations at the
MP2/aug-cc-pVTZ\cite{moller1934note} level of theory. For a given
geometry, the energy, forces and dipole moments are calculated using
the MOLPRO software package\cite{schuetz2018MOLPRO}. The data set not
only contained the structures of MA, AAA and AcAc, but included a
total of 49 structures and substructures, given in
Figure~S1, based on the ``amon''
approach.\cite{huang2017amons}\\

\noindent
An initial ensemble of 49\,000 molecular geometries was generated by
running Langevin dynamics at 1000 K with a time-step of $\Delta t =
0.1$ fs using the Atomic Simulation Environment
(ASE)\cite{larsen2017atomic} and the semi-empirical PM7
method\cite{stewart2007semiempirical} as implemented in
MOPAC.\cite{MOPAC2016stewar} For all these structures, reference data
at the MP2/aug-cc-pVTZ level of theory was computed and initial NNs
were trained on it. Based on the initial training, the data set was
augmented using adaptive
sampling.\cite{behler2016perspective,behler2015constructing} For this,
Langevin dynamics are run with two NNs and the initial data set is
extended with \textit{ab initio} data if the energies predicted by the
two NNs differed by more than 0.5 kcal/mol. Two rounds of adaptive
sampling were carried out.\\

\noindent
The final data set used to train the NNs contained energies, forces
and molecular dipole moments for 71\,208 structures including all
three molecules and their amons. Choosing the MP2/aug-cc-pVTZ level of
theory was motivated by the fact that it still provides good accuracy
but is also computationally feasible for data sets of size $\sim
10^5$. As the number of required training points is {\it a priori}
unknown when training a new NN the level of theory for the reference
calculations should be chosen with circumspection. This is why a
higher level method, such as CCSD(T), was not considered from the
outset but rather used in TL.\\

\noindent
{\it Transfer Learning to a Higher Level of Theory:} In order to
further improve the quality of the entire PES, TL \cite{pan2009survey}
was used. In TL, instead of starting the training procedure of an NN
from scratch, an NN trained on related data is used to initialize the
parameters, which usually leads to a better model with less data. This
approach can be useful when high-level reference data is scarce or
expensive to generate, whereas data for ``pre-training'' (in the
present case based on a lower level of electronic structure theory)
the NN is readily available.\cite{pan2009survey} As such, TL is a
valuable tool to bypass the high computational cost of modern
electronic structure calculations: First, an NN is pre-trained on a
large number of low level \textit{ab initio} data and then re-trained
with a smaller number of high level \textit{ab initio} data to
slightly adjust its parameters.\cite{smith2018outsmarting} \\

\noindent
Here, the model trained on the MP2 data is the base model and the pair
natural orbital (PNO-)-LCCSD(T)-F12/cc-pVTZ-F12
method\cite{werner2018coupled-cluster,peterson2008basis} is the higher
level of theory. A new data set at this level of theory is generated
by extracting structures from MD simulations run with the NN trained
on the MP2 reference data. The new data set contains a total of
49\,000 structures, composed of 1000 structures per molecule in the
reference set, see Figure~S1. Then, the data set is
split into a training set of 44\,100 and a test set of 4900
structures.  TL is carried out with different, randomly chosen
structures in the training set with sizes of 100, 1000, 5000, 15\,000,
25\,000 and 40\,000 structures and all parameters of the NN were
allowed to change. Since the reference data set only contains
energies, the loss function only depends on those and the dependency
on the forces and dipole moments was removed by setting the
corresponding weights in the loss term to zero. For comparison, models
with the same six training sets were also trained from scratch.\\

\subsection{Molecular Dynamics Simulations}
Molecular dynamics simulations are run using the NN-trained PESs to
determine the IR spectrum and the rate for HT. All MD simulations,
unless stated otherwise, were run as follows. Starting from an initial
structure, random momenta drawn from a Boltzmann distribution at 300 K
are used as initial conditions. The system is then equilibrated for 50
ps in the $NVT$ ensemble followed by 50 ps in the $NVE$ ensemble. This
was followed by MD simulation in the $NVE$ ensemble for a total of 1
ns, with a time step of 0.5 fs. For each of the molecules (MA, AAA,
and AcAc) a total of 1000 trajectories are run. Every second snapshot
along the trajectory is recorded for analysis.\\

\noindent
IR spectra are calculated from the dipole-dipole autocorrelation
function of the MD simulations. For every configuration the molecular
dipole moment $\boldsymbol{\mu} (t)$ is calculated following
Eq.~\ref{eq:physnet_dipole_moment}. From this, the dipole moment
autocorrelation function $C(t) = \left< \boldsymbol{\mu} (0)
\boldsymbol{\mu} (t) \right>$ is determined. A fast Fourier transform
of $C(t)$ yields $C(\omega)$ and a Blackman filter is employed to
minimize noise. Finally, the IR-spectrum is Boltzmann averaged to
yield
\begin{equation}
    A(\omega) = \omega\left[ 1- \exp(-\hbar \omega / (k_B T)) \right] C(\omega)
\end{equation}
Here, $\omega$ is the frequency, $\hbar$ is reduced Planck's constant,
$T$ is the temperature, and $k_B$ is the Boltzmann constant.\\

\noindent
HT rates are calculated from analyzing ``hazards'' which are deduced
from the residence times $t_k$ of the $k$-th
transition.\cite{helfand1978brownian,meuwly2002simulation} Here, the
residence time is defined as the time interval the transferring
hydrogen remains bound to one oxygen atom before changing to the other
oxygen. This can be measured by defining a cutoff O-H separation,
$r_c$. Whenever $r > r_c$, one transfer is considered completed and
the transition time $t_k$ is recorded. The Hazard for the $k$-th
transition\cite{helfand1978brownian}
\begin{align}
    H_k = \sum_i^{k-1} \frac{1}{n-i}
\end{align}
is determined from the $n$ residence times, $t_1$,$t_2$,...,$t_n$,
arranged in ascending order. From this, the rate can then be deduced
from the slope of the hazard plot ($H_k$ versus
$t_k$).\cite{helfand1978brownian}\\

\section{Results and Discussion}
First, the quality of the fully-dimensional, reactive PESs for the
three systems is discussed. This is followed by the analysis of the
computed IR spectra and the HT rates. Then, possibilities to exploit
chemical principles to reduce the number of required reference
structures and that for quality improvement based on
TL are examined.\\

\subsection{Quality of the PESs}
The performance of the NN is first validated on a separate test set
which includes structures randomly chosen from all molecules and their
amons. For the test set (9208 structures) a Pearson correlation
coefficient of $1-2.2\cdot10^{-7}$, a mean absolute error (MAE) in
energy of 0.020 kcal/mol and a root mean squared error (RMSE) for the
energy of 0.21 kcal/mol are obtained. The NN-predicted energies and
those from the MP2 reference data are compared in Figure~S2.
To test the reproducibility of these
results, an independent second NN model was trained with a MAE and
RMSE of 0.024 and 0.32 kcal/mol, respectively, close to the results
for the first NN.\\

\begin{figure}[htbp]
\centering
\includegraphics[width=0.7\textwidth]{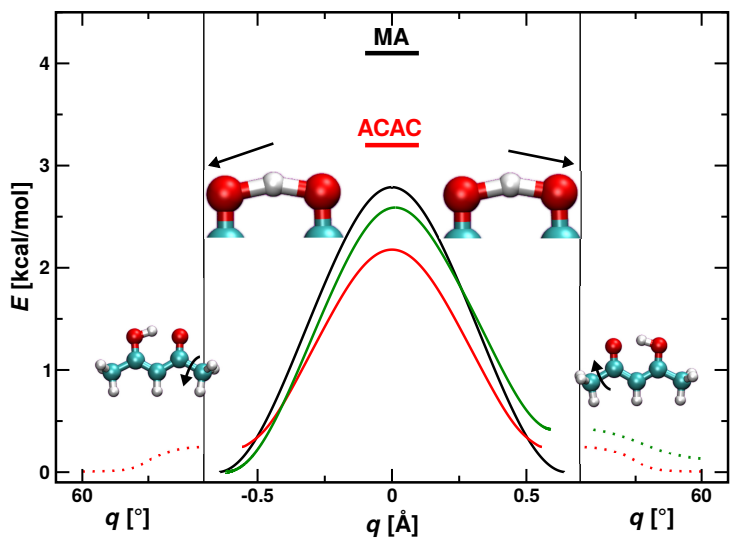}
\caption{The MEP from the NN-trained model for MA (black), AAA
  (green), and AcAc (red). The solid part of the lines corresponds to
  translation of the hydrogen along the reaction coordinate $q =
  r_{\rm O1H} -r_{\rm O2H}$, whereas the dotted part is for the
  rotation of a methyl group by $60^{\circ}$ (see insets). The short
  horizontal lines indicate the barrier heights at higher levels of
  theory.\cite{bowman2008malonaldehyde,howard2015infrared} The barrier
  heights on the NN PESs trained on the data from MP2/aug-cc-pVTZ
  calculations for MA, AAA, and AcAc are 2.79, 2.59, and 2.17
  kcal/mol, compared with higher-level calculations for MA (4.10
  kcal/mol at frozen-core CCSD(T)/(aug-)cc-pVTZ level whereby only the
  basis set on the oxygen atoms was
  augmented)\cite{bowman2008malonaldehyde}, and AcAc (3.20 kcal/mol at
  the CCSD(T)/cc-pVTZ level)\cite{howard2015infrared}.}
\label{fig:energy_profile}
\end{figure}

\noindent
The three energy profiles along the minimum energy path (MEP)
determined from the final NN-trained PESs are reported in Figure
\ref{fig:energy_profile}. The barrier heights predicted by the NN
increase from 2.17 kcal/mol for AcAc, to 2.59 kcal/mol for AAA and to
2.79 kcal/mol for MA, and compare with 2.18, 2.56, and 2.74 kcal/mol
from the reference MP2 calculations. Compared with high-level
calculations at the CCSD(T)/(aug-)cc-pVTZ level of theory, the barrier
heights for MA\cite{bowman2008malonaldehyde} and
AcAc\cite{howard2015infrared} are underestimated by somewhat more than
1 kcal/mol, though. Further improvements of the barriers from the NNs
can be achieved through TL, as will be discussed further below.\\

\begin{figure}[htbp]
\centering
\includegraphics[width=0.8\textwidth]{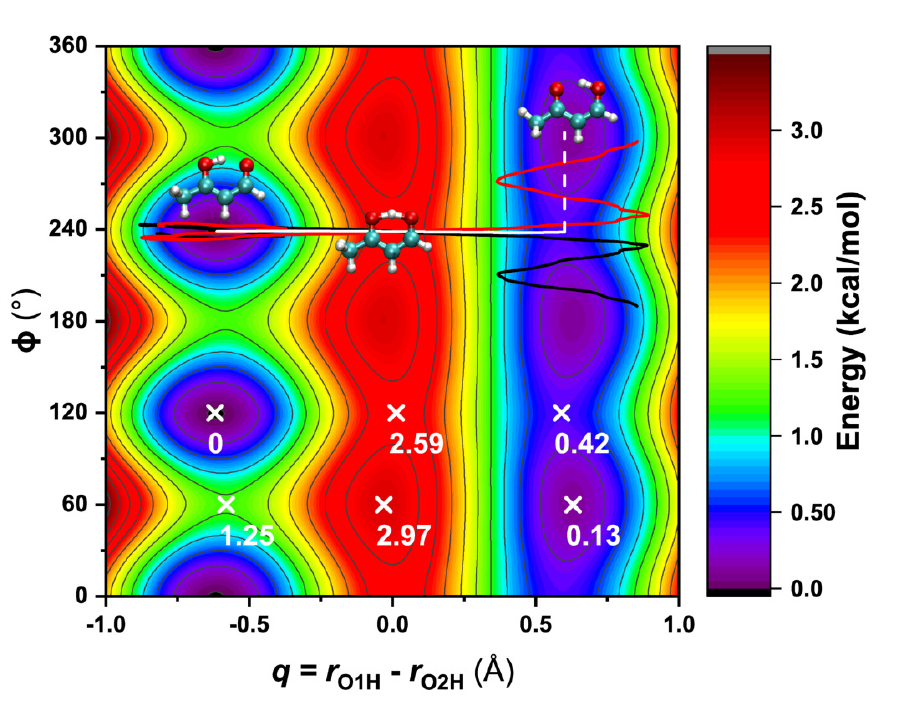}
\caption{Projection of the PES for AAA with $q = r_{\rm O1H} -r_{\rm
    O2H}$ and the methyl rotation as the two coordinates. Contour
  lines are drawn every 0.3~kcal/mol and important structures are
  shown in ball-and-stick representation. The white solid line is the
  part of the MEP for the HT and the dashed white line that for the
  rotation of the methyl group (see also
  Figure~\ref{fig:energy_profile}). The white crosses represent
  important extrema of the PES with the corresponding relative
  energies given in kcal/mol. The red and black lines are averages
  over 300 independent trajectories for two possible rotation
  directions of the methyl group. Important features on the PES are
  labelled and energies are reported relative to the global minimum
  (in kcal/mol).}
\label{fig:pes_met_malon}
\end{figure}

\noindent
For AAA a 2D-projection of the PES along the HT coordinate $q = r_{\rm
  O1H} -r_{\rm O2H}$ ($x-$axis) and the methyl rotation ($y-$axis) is
shown in Figure~\ref{fig:pes_met_malon}. As expected, a 3-fold
periodic pattern is found for the minima and the transition states
along the rotation of the methyl group by $120^{\circ}$ irrespective
of the value of $q$. With the hydrogen on the methylated side the
minima are shifted by $60^{\circ}$ compared with structures that have
the hydrogen on the unmethylated side. Upon crossing the TS, a
trajectory started from the isomer with the hydrogen atom on the
methylated side ends up in the conformation of the second isomer for
which the methyl is rotated by $60^{\circ}$ relative to the starting
structure.\\

\begin{figure}[htbp]
\centering
\includegraphics[width=0.6\textwidth]{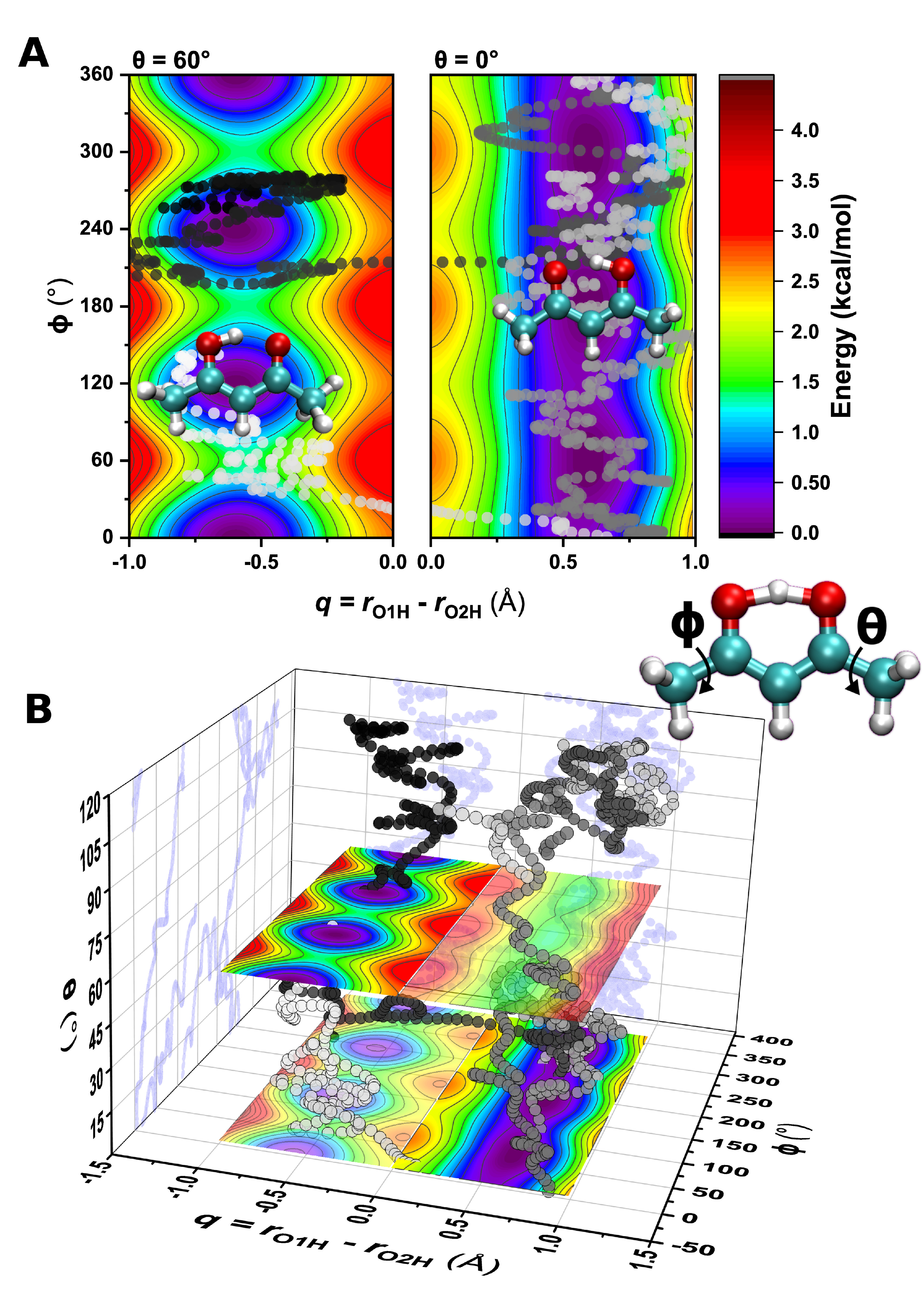}
\caption{ \textbf{A:} 2D projections of the AcAc PES with $q = r_{\rm
    O1H} - r_{\rm O2H}$ and $\Phi$ as the two coordinates. The
  contours are drawn every 0.3 kcal/mol with purple and red
  corresponding to low and high energies, respectively. They show the
  most extreme cases of the PES projections for $\Theta = 60^{\circ}$
  ($\Theta = 0^{\circ}$) for which an H--atom of the right methyl
  group is in an eclipsed (staggered) geometry with the adjacent
  O--atom (see insets). Pairs of ($q, \Phi$)--values are extracted
  from the first picosecond of a trajectory run at 300 K and are shown
  as dots, where black dots are the beginning and white dots mark the
  end of the trajectory. The complete projections of the surfaces are
  shown in panel B where the missing parts are illustrated as
  transparent. \textbf{B:} 3D representation of the same trajectory
  depending on $q=r_{\rm O1H} - r_{\rm O2H}$, $\Phi$ and $\Theta$ (see
  inset) with its $\Theta$-values illustrated modulo
  $120^{\circ}$. The two PES projections including the global minimum
  ($\Theta = 60^{\circ}$ and $\Theta = 0^{\circ}$) are included and
  repeat every $\Theta = 120^{\circ}$. The contours are shown every
  0.3 kcal/mol and have the same energy scale as
  Fig.~\ref{fig:pes_acac}A. The cluster of points in the
  upper right corner of the illustration (i.e. $q\approx0.5$~\r{A},
  $\Phi\approx300^{\circ}$ and $\Theta\approx105^{\circ}$) correspond
  to structures close to the global minimum and the clustering is
  further amplified because ($\Theta$ mod $120^\circ$) is used as the
  angle.}
\label{fig:pes_acac}
\end{figure}

\noindent
The two projections of the AcAc-PES containing the global minimum of
the molecule are partially shown in
Figure~\ref{fig:pes_acac}A with $q = r_{\rm O1H} -r_{\rm
  O2H}$ and the rotation angle $\Phi$ of the left methyl group as the
coordinates. In these projections the right-hand methyl group is
frozen in an eclipsed ($\Theta = 60^{\circ}$) or in a staggered
($\Theta = 0^{\circ}$) geometry relative to the adjacent O--atom.  The
PESs shown here are again periodic (periodicity of $120^{\circ}$) with
respect to the rotation angle $\Phi$ of the methyl group. For $q \approx
-0.6$ \r{A} the rotation of the left methyl group involves a barrier
of 1.37 kcal/mol which reduces to 0.25 kcal/mol for $q \approx
0.6$~\r{A}. Finally, for $q < -0.9$ \r{A} or $q > 0.9$ \r{A}
(i.e. along the O--O separation) the potential energy increases.\\
  
\noindent
For visualization purposes, $q$, $\Phi$ and $\Theta$ are extracted
from a trajectory used for the computation of the IR spectrum and the
first 1 ps is shown in Figure~\ref{fig:pes_acac}A as well as in a 3D
representation in Figure~\ref{fig:pes_acac}B. Black dots represent
early and white dots represent late points during the trajectory. The
hydrogen atom is transferred twice and the left-hand side methyl
rotates due to the lower rotational barrier for $q>0$ (see
Fig.~\ref{fig:pes_acac}A). When interpreting the trajectory it should,
however, be taken into account that during the MD simulations both
methyl groups are free to rotate whereas for the PES scan the
right-hand methyl is frozen.  The 3D trajectory illustrates that both
methyl rotations (along the $y-$ and $z-$axis) and HT (along the
$x-$axis) occur during the course of an equilibrium dynamics and that
they are coupled as the barrier for methyl rotation depends on whether
the oxygen atom two bonds away from the methyl group carries the
hydrogen or not.\\

\begin{figure}[htbp]
\centering
\includegraphics[width=0.8\textwidth]{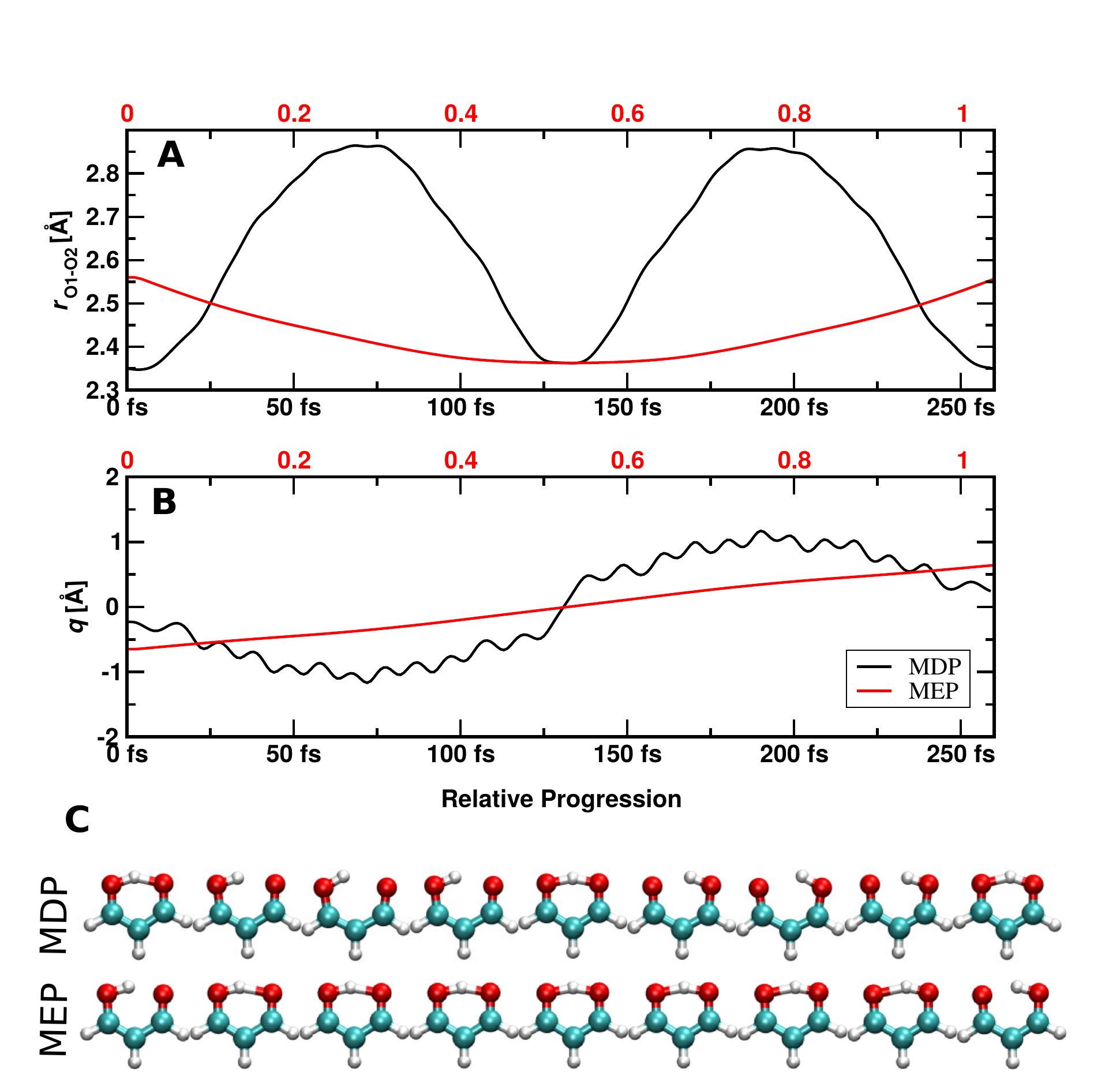}
\caption{The MEP and MDP for MA. The O1--O2 distance $r_{\rm O1-O2}$
  (\textbf{A}) and the reaction coordinate $q = r_{\rm O1H}
  -r_{\rm O2H}$ (\textbf{B}) for the MDP (black) and the MEP (red) are
  shown. The MEP lacks important dynamical information such as the
  oscillation of the O1--O2 distance (\textbf{A}) or the oscillation of
  the O-H bond (\textbf{B}). These aspects present in the MDP
  play an important role for promoting the reaction. The snapshots
  shown in (\textbf{C}) illustrate the atomic displacements for
  the two paths. The \textit{x}-axis for the MEP (red) corresponds to
  the transfer of the H--atom progressing from the initial minimum
  energy structure (0) to the TS (0.5) and to the final optimum
  structure with the H--atom completely transferred (1). Conversely,
  the MDP progresses in time (black \textit{x}-axis).}
\label{fig:mepvsmdpmalonaldehyde}
\end{figure}

\noindent
Next, MEP and minimum dynamical path (MDP) for MA are compared. The
MDP is defined as the path followed by a trajectory that passes the
exact transition state with zero excess energy.\cite{MM.mdp:2019} The
MEP for MA is symmetrical (see Figure~\ref{fig:mepvsmdpmalonaldehyde})
with a transition state at $E_{\rm TS} = 2.79$ kcal/mol above the
energy minimum. For both, MEP and MDP the O1--O2 distance and the
reaction coordinate $q = r_{\rm O1H} -r_{\rm O2H}$ are
considered. Moreover, snapshots at fixed intervals of the two paths
are illustrated. Although the snapshots of the MDP and MEP are
similar, the MEP lacks important dynamical information: First, the MDP
reveals a strong oscillation of the O--O distance (see $r_{\rm O1-O2}$
in Figure~\ref{fig:mepvsmdpmalonaldehyde}), i.e. the O1--O2 distance
is coupled to the HT coordinate. Even though both MEP and MDP indicate
that the two oxygen atoms approach each other during HT, the strong
counter-movement with O1--O2 distances up to 2.85 \r{A} only manifests
itself in the MDP. This could be essential in promoting HT and is not
reflected in the MEP. A similar effect is seen for the translation of
the hydrogen atom, see Figure~\ref{fig:mepvsmdpmalonaldehyde}B.\\

\noindent
For AAA the MEP is shown as a projection in
Figure~\ref{fig:pes_met_malon} (white solid and dotted trace). It
demonstrates that hydrogen translation along $q$ is not coupled to the
methyl rotation. To obtain insight into the dynamics of a
representative trajectory, MD simulations are run from structures
close to the TS with an energy of $E = E_{\rm TS} + 3$ kcal/mol,
following the approach presented in
Ref.~\citenum{MM.mdp:2019}. Overall, 300 independent trajectories were
run and fall into two groups, depending on the direction of the
rotation of the methyl group after HT. Between $q \approx -0.5$ to $q
\approx 0.5$ \r{A} the MEP (white) and the path traced by the
trajectories (red and black traces) follow comparable paths across the
barrier. Once they reach the TS around $q = 0.6$ \AA\/ the coupling
between HT and methyl rotation leads to differences between MEP and
the averaged trajectories. However, this coupling is rather dynamical
in nature and not conveyed by the shape of the PES. Due to the finite
kinetic energy the motion continues beyond the TS with energy 0.42
kcal/mol above the minimum into regions with $q>0.6$ where coupling
between $q$ and $\Phi$ sets in. Coupling of HT and methyl rotation is
also consistent with results from neutron scattering
experiments.\cite{trommsdorff:2002}\\

\subsection{Infrared Spectroscopy}
All IR spectra shown in this work are calculated from NN-MD
simulations. The IR spectrum for MA, AAA and AcAc are discussed in the
following.

\begin{figure}[htbp]
\centering
\includegraphics[width=0.9\textwidth]{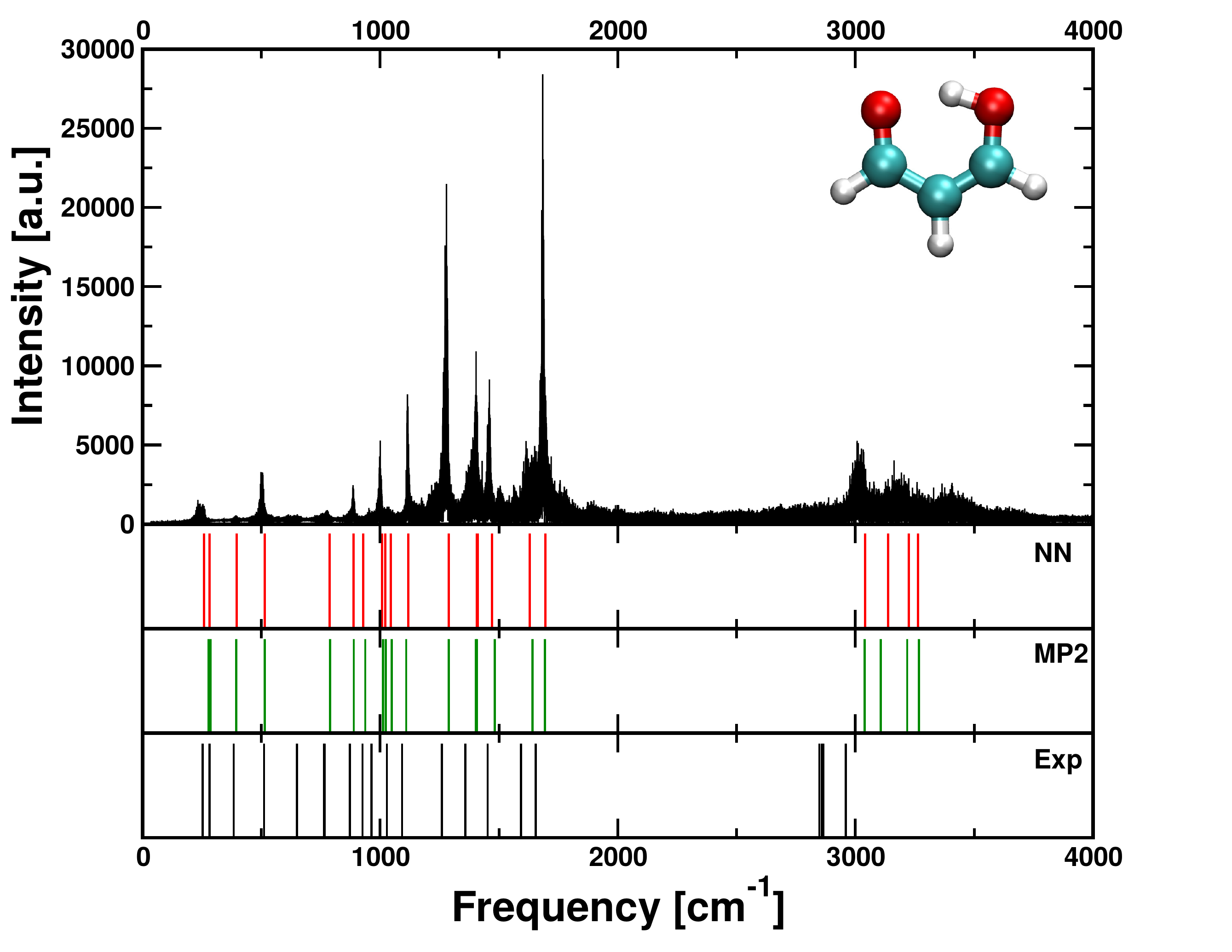}
\caption{IR spectrum of MA at 300 K averaged over 1000 independent
  runs, each 1 ns in length (top panel). The normal modes determined
  from the NN PES are given in the lower panel (red) together with
  normal modes determined from MP2 reference calculations (green) and
  the center frequencies from experiment (black).\cite{maexp83} HT
  leads to broadening of the spectrum between 2000 and 3000
  cm$^{-1}$.\cite{howard2015infrared} The experimental spectrum shows
  a peak at 650~cm$^{-1}$ which is not covered by the theoretical
  normal modes, however, its presence can be guessed in the
  IR-spectrum. The peak at 650~cm$^{-1}$ is attributed to a ring bend,
  but the frequency was not clearly assigned to a fundamental
  vibration.\cite{howard2015infrared} Moreover, the position of the
  $\nu_{\rm CH}$ is at lower frequencies than predicted by the
  theoretical values.}
\label{fig:ir_malon}
\end{figure}

\noindent
\paragraph{Malonaldehyde (MA)} The
calculated IR spectrum from MD simulations at 300 K along with the
normal modes from the NN is shown in Figure~\ref{fig:ir_malon}. The
center frequencies of the IR spectrum show good overall agreement with
the normal modes, although single peaks are shifted due to
anharmonicity and couplings. Most importantly, the broad absorption
between 2000 and 3000 cm$^{-1}$ in the IR spectrum is due to the
transferring hydrogen. There are 149\,073 HT events in the 1000
independent trajectories, each 1 ns in length, from which the spectra
were determined. To examine the accuracy of the NN normal mode
predictions, they are compared with those computed directly from MP2
calculations, see Table \ref{tab:comparison_NM}. The largest deviation
is 8.3 cm$^{-1}$ for mode 11 and the RMSD is 3.6 cm$^{-1}$. The normal
modes from the MP2 calculations and the NN trained on them also
compare well with higher-level calculations using
CCSD(T)/(aug-)cc-pVTZ.\cite{bowman2008malonaldehyde} As a comparison
of the accuracy of the NN, the RMSD for the normal modes from the
CCSD(T) calculations and the fitted surface using permutationally
invariant polynomials based on them is 21.3 cm$^{-1}$ which compares
with 3.6 cm$^{-1}$ from the NN-trained PES.\\

\noindent
A detailed comparison and interpretation of the infrared spectrum with
that from experiment is outside the scope of the present
work. Instead, a few features are highlighted and the experimental
line positions for the fundamentals\cite{maexp83} are provided as a
guide in Figure~\ref{fig:ir_malon}. The C--H stretching vibrations
$\nu_{CH}$ are assigned to peaks in the 3000 cm$^{-1}$ region although
experimentally, they are around or even below 3000 cm$^{-1}$. However,
it is well known that harmonic frequencies from \textit{ab initio}
calculations need to be scaled when comparing them with experiments
\cite{scott1996harmonic}. For the MP2/aug-cc-pVTZ the scaling is 0.953
which brings the frequencies to values between 2897 and
3110~cm$^{-1}$. The broad absorption between 2000 and 3000 cm$^{-1}$
in the IR spectrum is due to the transferring hydrogen. This region is
spectroscopically empty when considering the normal modes. Around 1600
cm$^{-1}$ the C-O stretch vibrations follow which are at 1693 and 1640
cm$^{-1}$, respectively. At yet lower frequency, framework modes
follow which are highly mixed. Their assignment is outside the scope
of the present work. Finally, the O--O stretch vibration in the IR
spectrum is at 243~cm$^{-1}$ whereby the assignment was made by
analyzing the normal modes computed on the NN-trained PES.\\

\begin{table}[htbp]
\begin{tabular}{r||r|r||r|r}
\textbf{Mode} & \textbf{NN [cm$^{-1}$]}& \textbf{MP2 [cm$^{-1}$]} & \textbf{PES\cite{bowman2008malonaldehyde} [cm$^{-1}$]}    & 
\textbf{CCSD(T)\cite{bowman2008malonaldehyde}[cm$^{-1}$]}\\ \hline
1	&	277.8	&	277.5	&	268.6	&	252.9		\\
2	&	294.7	&	286.6	&	295.4	&	283.3		\\
3	&	393.8	&	394.3	&	383.2	&	373.8		\\
4	&	512.7	&	514.0	&	522.1	&	504.2		\\
5	&	787.9	&	789.4	&	760.6	&	780.6		\\
6	&	888.7	&	888.6	&	888.3	&	887.5		\\
7	&	937.6	&	937.6	&	897.4	&	899.1		\\
8	&	1016.9	&	1012.3	&	995.7	&	992.9		\\
9	&	1025.0	&	1023.8	&	998.0	&	1004.1		\\
10	&	1049.8	&	1048.7	&	1023.1	&	1032.3		\\
11	&	1118.0	&	1109.7	&	1105.4	&	1102.2		\\
12	&	1289.5	&	1288.3	&	1280.8	&	1276.2		\\
13	&	1405.1	&	1403.1	&	1393.6	&	1403.1		\\
14	&	1413.5	&	1408.0	&	1419.5	&	1409.9		\\
15	&	1478.6	&	1482.0	&	1490.2	&	1473.2		\\
16	&	1640.4	&	1641.5	&	1647.2	&	1636.0		\\
17	&	1692.8	&	1692.9	&	1713.5	&	1698.3		\\
18	&	3037.1	&	3039.0	&	3020.7	&	3009.1		\\
19	&	3114.4	&	3107.1	&	3196.8	&	3183.3		\\
20	&	3216.5	&	3217.9	&	3251.4	&	3236.9		\\
21	&	3269.1	&	3267.3	&	3348.9	&	3266.4		\\ \hline 
\textbf{RMSD} &  \multicolumn{2}{c||}{3.6 cm$^{-1}$} & 
\multicolumn{2}{c}{21.3 cm$^{-1}$}
\end{tabular}
\caption{The normal modes for MA calculated on the NN PES compared
  with those based on the reference calculations at the
  MP2/aug-cc-pVTZ level of theory. The RMSD between the two is 3.6
  cm$^{-1}$. As a comparison of both, the absolute wavenumbers and the
  quality of the fit of the PES results from normal modes at the
  CCSD(T)/(aug-)cc-pVTZ level of theory and the PES fitted to
  permutationally invariant polynomials are
  provided.\cite{bowman2008malonaldehyde} There, the average
  difference is 21.3 cm$^{-1}$.}
\label{tab:comparison_NM}
\end{table}

\begin{figure}[htbp]
\centering \includegraphics[width=0.9\textwidth]{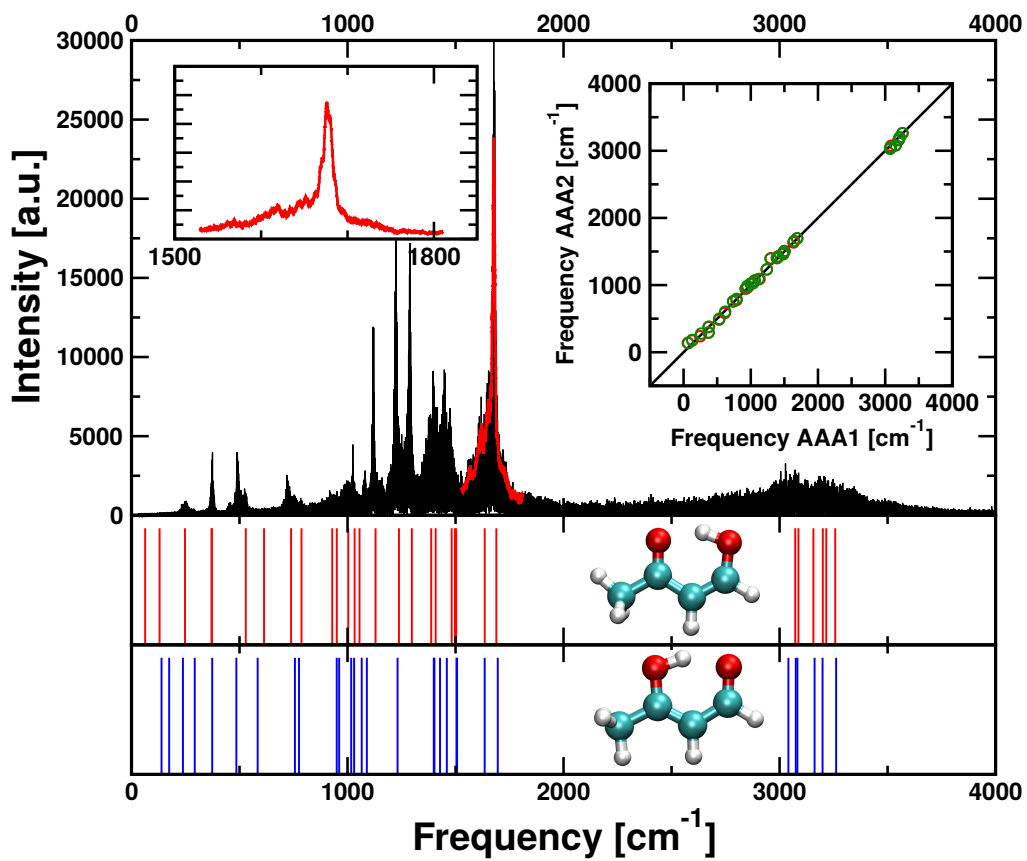}
\caption{IR spectrum of AAA at 300 K averaged over 1000 independent MD
  runs (top panel) together with the normal modes of the two isomers
  predicted by the NN. Some peaks of the IR spectrum are due to
  sampling both isomers. The inset in the upper right corner
  illustrates the correlation between the two sets of normal modes for
  both isomers from the NN (red) and MP2 calculations (green).}
\label{fig:ir_met_malon}
\end{figure}

\noindent
\paragraph{Acetoacetaldehyde (AAA)} The calculated IR spectrum for AAA from MD
simulations is reported in Figure~\ref{fig:ir_met_malon} together with
the normal modes for the two isomers. Depending on the position of the
transferring hydrogen the normal mode frequencies can change
appreciably. Because in the finite-temperature simulations both
isomers are sampled due to HT, the IR spectrum contains signatures
from both isomers. The normal mode frequencies of both isomers are
directly compared in the upper right inset in
Figure~\ref{fig:ir_met_malon} for the NN-predicted (red) and the MP2
(green) values. Although the two sets of frequencies correlate, slight
deviations from the diagonal occur which are due to differences in the
intramolecular interactions. Similar to MA, the O--O vibration is
assigned to the IR frequency of 249~cm$^{-1}$ by examining the normal
modes of the two isomers (261 and 277~cm$^{-1}$).\\

\begin{figure}[htbp]
\centering
\includegraphics[width=0.9\textwidth]{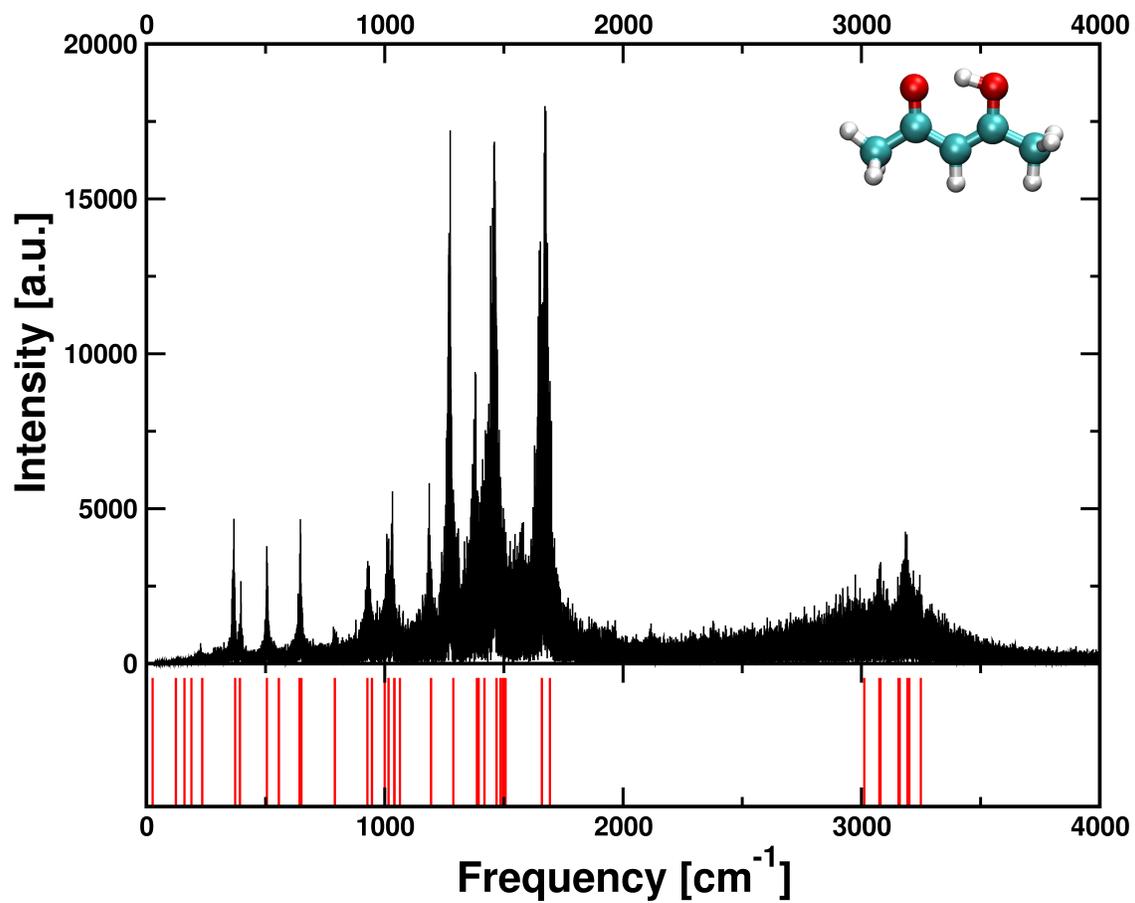}
\caption{IR spectrum of AcAc obtained from 1000 independent MD
  simulations run at 300~K on the NN PES (upper panel) compared with
  the normal modes calculated on the same PES (lower panel).}
\label{fig:ir_acac}
\end{figure}

\noindent
\paragraph{Acetylacetone (AcAc)} The calculated IR spectrum of AcAc is shown in
Figure~\ref{fig:ir_acac}, complemented with the normal modes predicted
by the NN. Again, the normal modes are well represented by the IR
spectrum, with slight shifts due to the anharmonicity. Moreover, good
agreement can be found with experiment and
theory\cite{howard2015infrared}. Some prominent peaks and regions are
assigned as follows: The broad band between 2000 and 3000 cm$^{-1}$ is
assigned to $\nu_{OH}$, which is caused by the transferring
hydrogen. The C-H vibrations are slightly higher in wavenumber and are
around and above 3000 cm$^{-1}$. Similar to MA, the CH stretch
vibrations in the experiment are rather around 3000 cm$^{-1}$ whereas
those from computations are shifted to the blue. The reason for this
is unclear. For the signatures in the 1000 to 2000 cm$^{-1}$ region
the agreement between the normal modes and the IR-spectrum is quite
apparent. At yet lower frequencies it is found that certain normal
modes do not appear in the infrared spectrum as the dipole moment
along these motions does not change and, hence, no IR-activity is
expected. The normal mode frequency of the O--O stretch vibration is
at 234 cm$^{-1}$ compared with the corresponding peak in the IR
spectrum at 226~cm$^{-1}$.\\

\subsection{Hydrogen Transfer Rates}
\begin{figure}[htbp]
\centering
\includegraphics[width=0.9\textwidth]{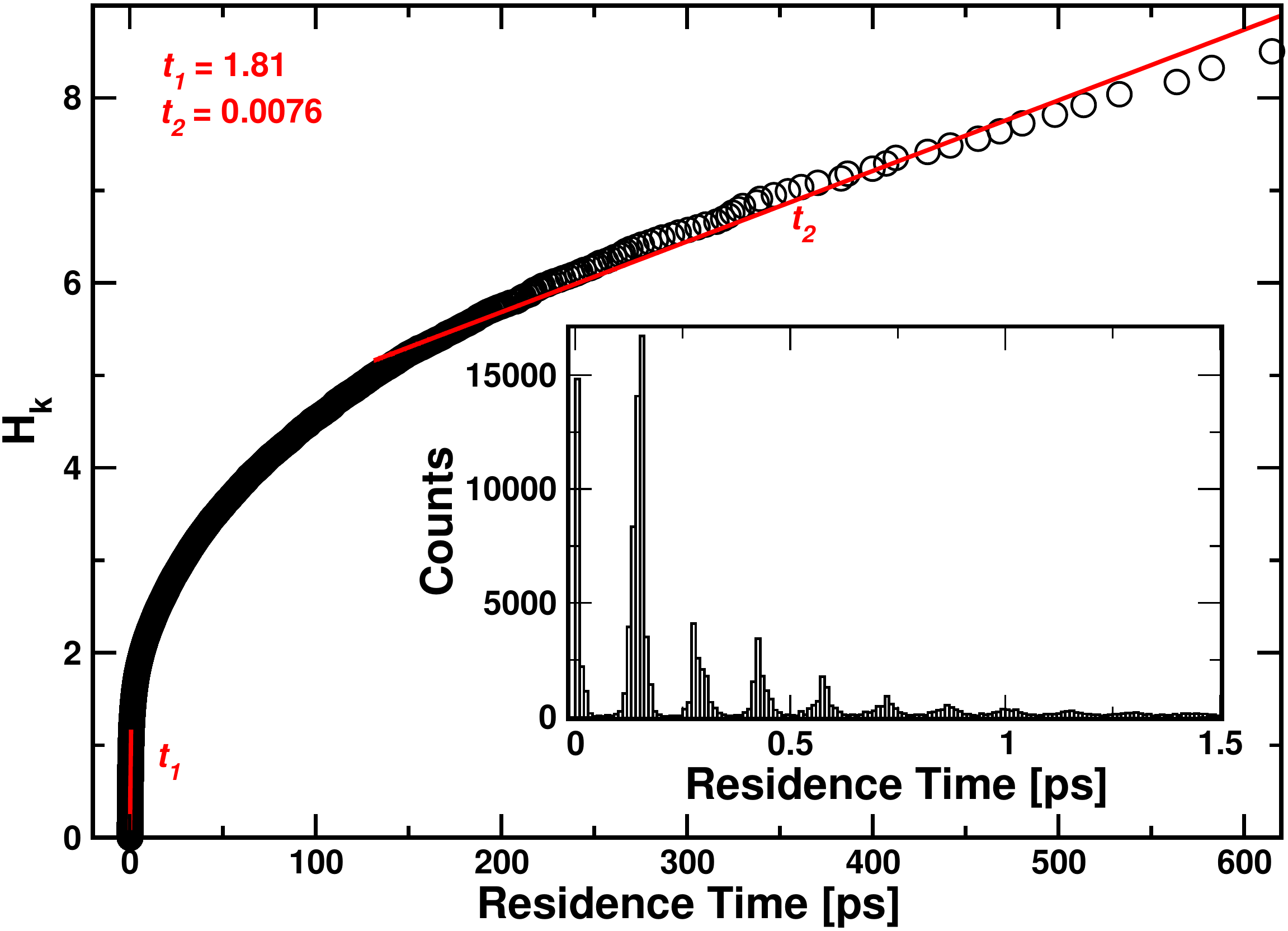}
\caption{Hazard plot for the residence times for a total of 1 $\mu$s
  simulation time of MA at 300~K (only every fifth data point is
  shown). About 65 \% of the transfers occur with residence times $<
  0.6$ ps and the longest residence time encountered was $\approx 970$
  ps. A linear regression for $0 < \tau < 0.6$ ps yields a rate of
  $t_1 = 1.81$ crossings/ps. For the long residence times a rate of
  $t_2 = 0.0076$ crossings/ps is found. The inset shows the
  distribution $p(\tau)$ of residence times, which exhibits a period
  of $\approx 0.14$ ps, corresponding to a frequency of 240 cm$^{-1}$
  which is close to the O--O stretch frequency at 243~cm$^{-1}$.}
\label{fig:ma_hazard}
\end{figure}

HT rates were calculated from a Hazard
analysis.\cite{helfand1978brownian,meuwly2002simulation} For MA the
analysis was first carried out with $r_c = 1.23$ \r{A} which is
slightly longer than the O--H separation in the transition state which
is 1.20 \r{A}. The Hazards $H_k$ are reported in
Figure~\ref{fig:ma_hazard} with the residence time distribution
$p(\tau)$ for the first 1.5 ps shown in the inset. The periodicity of
$p(\tau)$ is $\sim 0.14$ ps which corresponds to a frequency of
$7.14$/ps or a vibration of 240 cm$^{-1}$. This is close to the
frequency of the O--O stretching vibration at 243 cm$^{-1}$ and
suggests that the O--O motion is gating HT. A similar situation was
found in protonated ammonia dimer for which the N--N vibration was
also found to be the frequency to which HT is
coupled.\cite{meuwly2002simulation}.\\

\noindent
The fast reaction rate for HT in MA is $t_1 = 1.81$ crossings/ps and
the slow process has a time constant $t_2 = 0.0076$ crossings/ps. The
short time scale corresponds to ``ballistic'' transfer whereas the
long time scale is a ``dwell'' time scale caused by partial
equilibration of the hydrogen in one of the two potential wells. In
order to test whether the findings depend on the choice of $r_c$, the
same data was also analyzed using $r_c = 1.1$ \r A (see
Figure~S3). With this, the fast rate reduces
slightly from $t_1 = 1.81$ to $t_1 = 1.60$ crossings/ps. This is
explained by the exclusion of a set of HTs which immediately
re-transfer without leaving the TS region completely. The slow
reaction rate was not influenced by the change in $r_c$.\\

\noindent
On the other hand, $t_2$ depends quite sensitively on the range over
which the data is fitted for long residence times. Values ranging from
0.006 to 0.008 crossings/ps are obtained depending on the range over
which the fit is carried out. Within transition state theory (and
using a frequency factor of $10^{12}$
s$^{-1}$)\cite{anslyn2006modern}, a transition rate of 0.0076
ps$^{-1}$ corresponds to a barrier height of $E_b = 2.91$ kcal/mol,
consistent with 2.79 kcal/mol from the NN PES, see
Figure~\ref{fig:energy_profile}. Finally, in previous work, the
hopping rate for MA was determined to be 0.0024 crossings/ps from
simulations using a scaled MP2 PES with a barrier of 4.3 kcal/mol,
consistent with higher-level CCSD(T)
calculations.\cite{yang2010generalized} This compares with an
increased rate for HT (0.0076 crossings/ps) on the MP2 NN-PES due to
its lower barrier height.\\

\noindent
The same analysis was carried out for AAA and AcAc using $r_c = 1.23$
\AA\/, see
Figures~S4~and~S5. Due to
the smaller barrier heights for HT it is expected that some of the
rates increase compared with those in MA. For AAA, the fast and slow
rates are $t_1 = 1.56$ crossings/ps and $t_2 = 0.015$
crossings/ps. Similar to MA, the distribution of the transition times
$p(\tau)$ shows a periodicity of $\approx 0.135$ ps, which corresponds
to a frequency of 247 cm$^{-1}$ which is again close to the frequency
for the O--O motion at 249~cm$^{-1}$. For AcAc the fast and slow rates
are $t_1 = 1.45$ crossings/ps and $t_2 = 0.036$ crossings/ps and the
periodicity of $p(\tau)$ yields a frequency of 222 cm$^{-1}$. This is
in good agreement with the IR frequency at 226~cm$^{-1}$. In
conclusion, the fast time scale for all three systems is between 1.5
and 1.8 crossings/ps whereas the long time scale $t_2$ differs
appreciably and ranges from 0.008 crossings/ps to 0.036 crossings/ps,
commensurate with a difference in the barrier height by almost a
factor of two between MA and AcAc.\\

\noindent
An independent test for the applicability of a Hazard analysis is to
consider the total number of hazards as a starting population and to
follow its decay using first order kinetics. Then, the residence time
is the time for a ``state'' to decay. For MA such an analysis yields
1.69 crossings/ps compared with 1.81 crossings/ps from the Hazard
analysis and confirms its validity. In all these considerations it is
important to note that this approach using classical trajectories is
expected to give a lower limit for the rate because quantum effects,
such as tunneling or zero-point energy are not included and typically
increase the rate.\\

\subsection{Generalizability of the NN}
High-quality PESs are often difficult to obtain due to the
computational cost in calculating a sufficiently large number of
\textit{ab initio} reference energies at a sufficiently high level of
theory. It would be desirable to minimize the number of such
high-level calculations that need to be carried out while retaining
the quality of the high-level calculations in the represented PES. Two
possibilities are explored in the following. The first examines
whether it is possible to infer the PES for a larger chemical compound
from information about a smaller one (transferability). The second one
attempts to learn a high-level PES from a fully-dimensional PES at a
lower level of theory together with information about selected points
of the higher-level PES (TL).\\

\begin{figure}[htbp]
\centering
\includegraphics[width=0.5\textwidth]{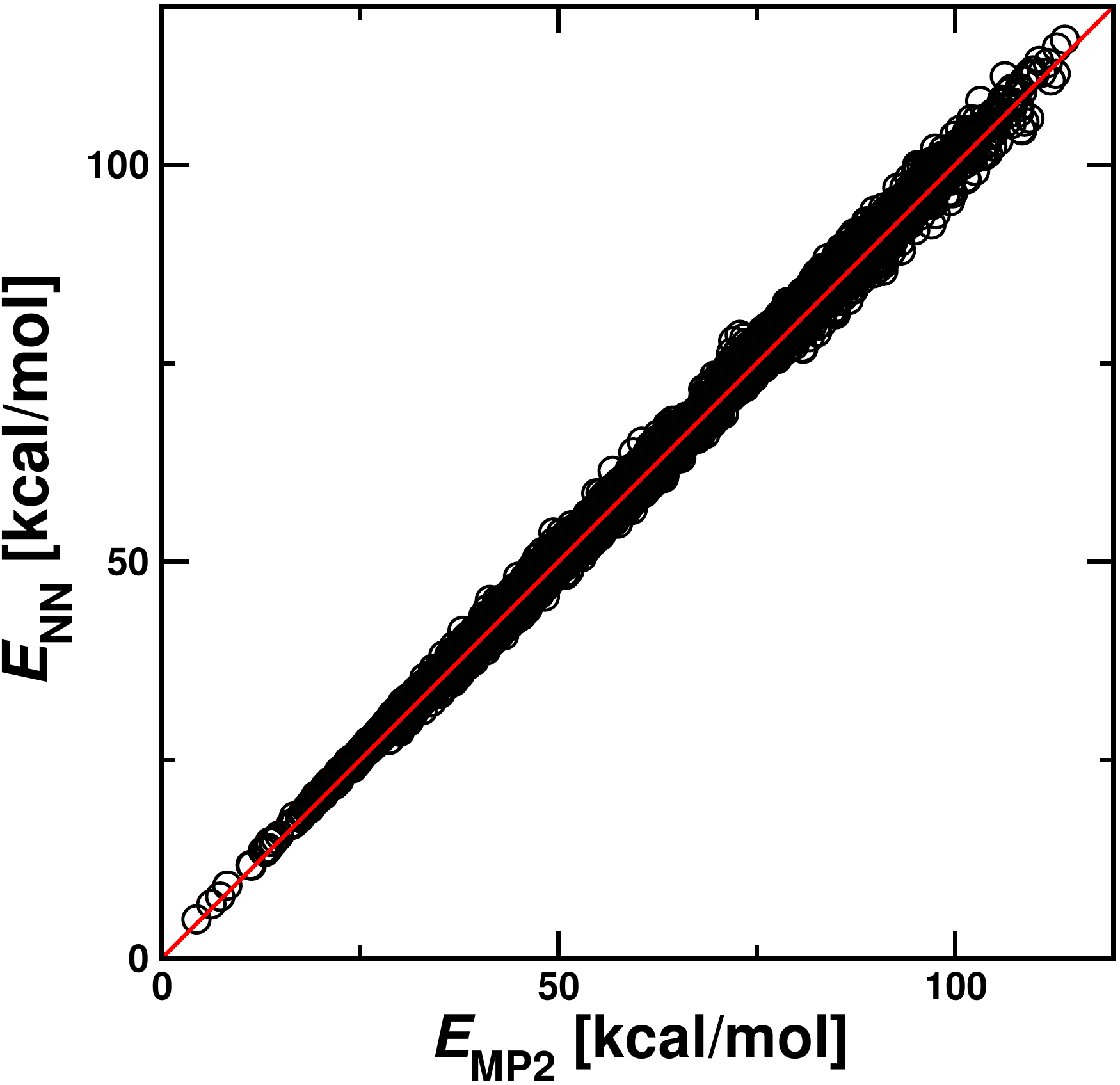}
\caption{Performance of NN* from analyzing 5000 AcAc structures from
  training on MA, AAA, and amons. The energies predicted by NN* are
  compared with those from the reference MP2/aug-cc-pVTZ calculations
  and demonstrate that the H$\rightarrow$CH$_3$ replacement can be
  predicted with ``chemical accuracy''.  The MAE is 0.9775~kcal/mol,
  and NN* achieves a Pearson correlation coefficient of
  $r^2=0.9984$. The zero of energy is the optimized AcAc structure.}
\label{fig:generalizability}
\end{figure}

\noindent
To assess how the NN generalizes on unseen structures, an NN PES for
AcAc (two CH$_3$) is determined by learning on a training set with
structures for MA (no CH$_3$), AAA (one CH$_3$) and the amons
only. This model is called NN* in the following. The performance of
NN* is reported in Figure~\ref{fig:generalizability} where the unknown
energies of 5000 AcAc structures are predicted from NN* and compared
with the reference energies at the MP2/aug-cc-pVTZ level of
theory. The Pearson correlation coefficient is $r^2=0.9984$, the
mean absolute error (MAE) is 0.98 and the root mean squared error
(RMSE) is 1.25 kcal/mol. The data shows only a small, uniform
dispersion without evident outliers. It is noteworthy that the MAE is
below 1 kcal/mol, i.e. ``chemical accuracy'' is achieved for this
H$\rightarrow$CH$_3$ replacement over a range of energies exceeding
100 kcal/mol.\\

\begin{figure}[htbp]
\centering
\includegraphics[width=0.8\textwidth]{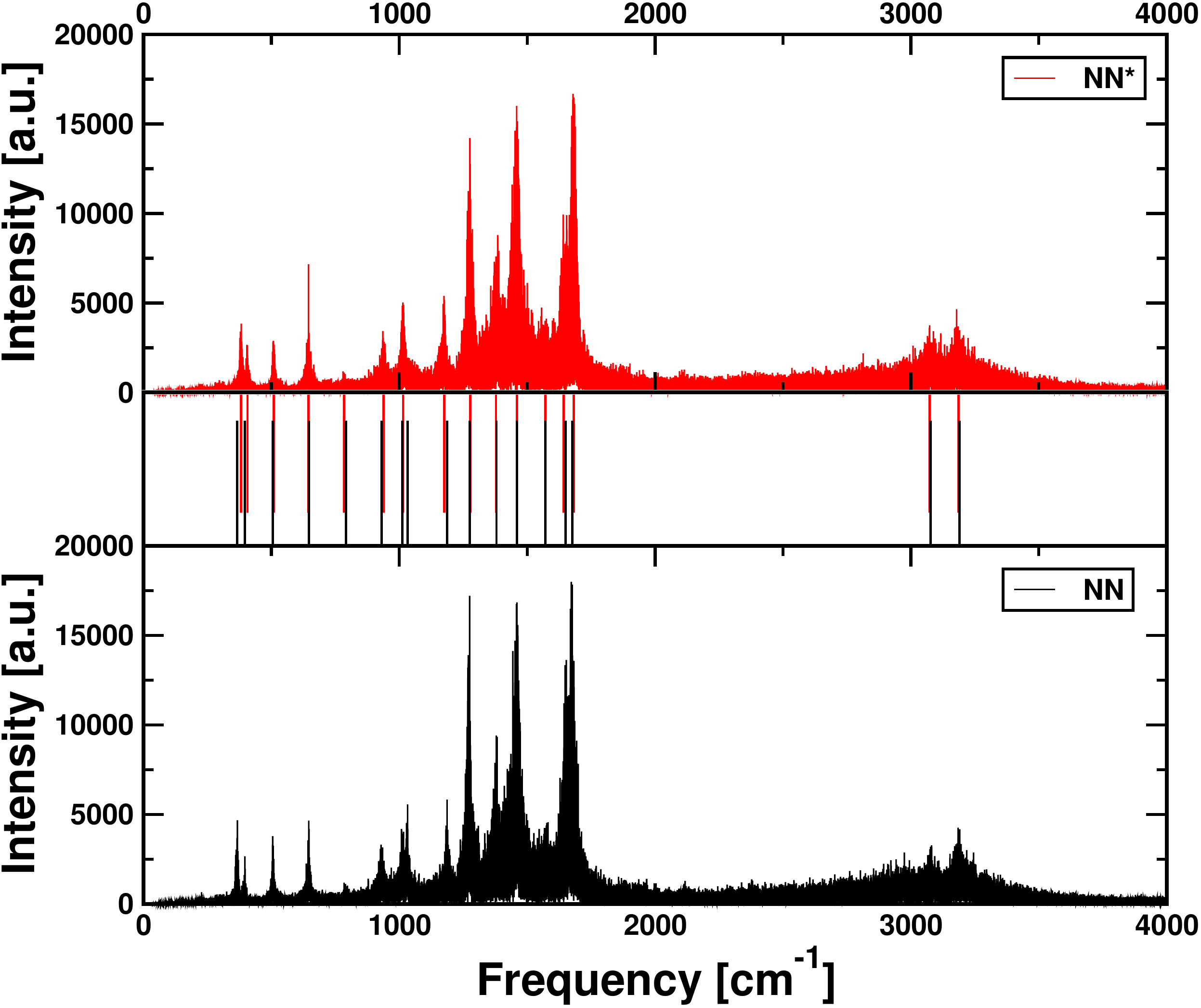}
\caption{Comparison of the IR-spectra from MDs for AcAc from models
  NN* (top panel, red) and from NN (bottom panel, black). The red and
  black impulse lines in the middle panel show the positions of the most
  prominent peaks. They differ by at most 14.6 cm$^{-1}$.}
\label{fig:ir_comparison}
\end{figure}

\noindent
The performance of NN* is further examined by computing the IR
spectrum and comparing it with that from the NN PES trained on the
data set including AcAc. The results are summarized in
Figure~\ref{fig:ir_comparison}, where the IR spectrum from using the
NN PES is shown in the bottom and the spectrum from using NN* in the
top panel. The stick spectrum in the middle panel are the peak
positions of the IR spectrum in the corresponding colors. The IR
spectra calculated from NN and NN* are very similar with a maximum
difference of the peak positions of 14.6 cm$^{-1}$ for the line at
366.6 (381.2) cm$^{-1}$. The average difference is, however, only 5.4
cm$^{-1}$ (see Tab.~\ref{tab:comparison_acac_ir_peaks}). Moreover, one
additional peak was found in the IR-spectrum calculated by the final
model.
A normal mode analysis for NN and NN* shows that all the major
differences are caused by vibrations involving the additional methyl
group, see Figure~S6.\\

\noindent
The results discussed above show that an NN can generalize to unknown
structures for an H$\rightarrow$CH$_3$ substitution with an error of
$<1$ kcal/mol for energies and for normal modes with a mean average
error of 5.4 cm$^{-1}$. Hence, PhysNet appears to be a suitable model
to further explore the possibility to generalize fully dimensional,
reactive PESs based on smaller reference structures which may increase
the speed with which such models can be trained.\\

\begin{table}[htbp]
\begin{tabular}{c|c|c}
\textbf{NN* [cm$^{-1}$]} & \textbf{NN 
[cm$^{-1}$]} & \textbf{Absolute Difference [cm$^{-1}$]}\\ \hline
 381.2  & 366.6 & 14.6\\
407.0&  396.7 & 10.3\\
510.1 &  506.6 & 3.5\\
644.8 &  646.4 & 1.6\\
784.7 &  791.5 & 6.8\\
939.4 &  930.9 & 8.5\\
1015.0 &  1011.4 & 3.6\\
- &  1032.4 & -\\
 1176.0 &  1187.1 & 11.1\\
1277.7 &  1275.7 & 2.0\\
1378.0 &  1379.9 & 1.9\\
1459.7 &  1460.5 & 0.8\\
1571.7 &  1571.6 & 0.1\\
1642.6 &  1650.4 & 7.8\\
1682.4 &  1677.0 & 5.4\\
3072.9  & 3077.3 & 4.4\\
3185.9 & 3189.2 & 3.3\\ \hline
& & \textbf{average = 5.36 cm$^{-1}$}
\end{tabular}
\caption{Comparison of the most prominent IR peaks for AcAc from NN
  and NN*, see Figure~\ref{fig:ir_comparison}.}
\label{tab:comparison_acac_ir_peaks}
\end{table}

\subsection{Transfer Learning to a Higher Level of Theory}
\begin{figure}[htbp]
\centering
\includegraphics[width=0.8\textwidth]{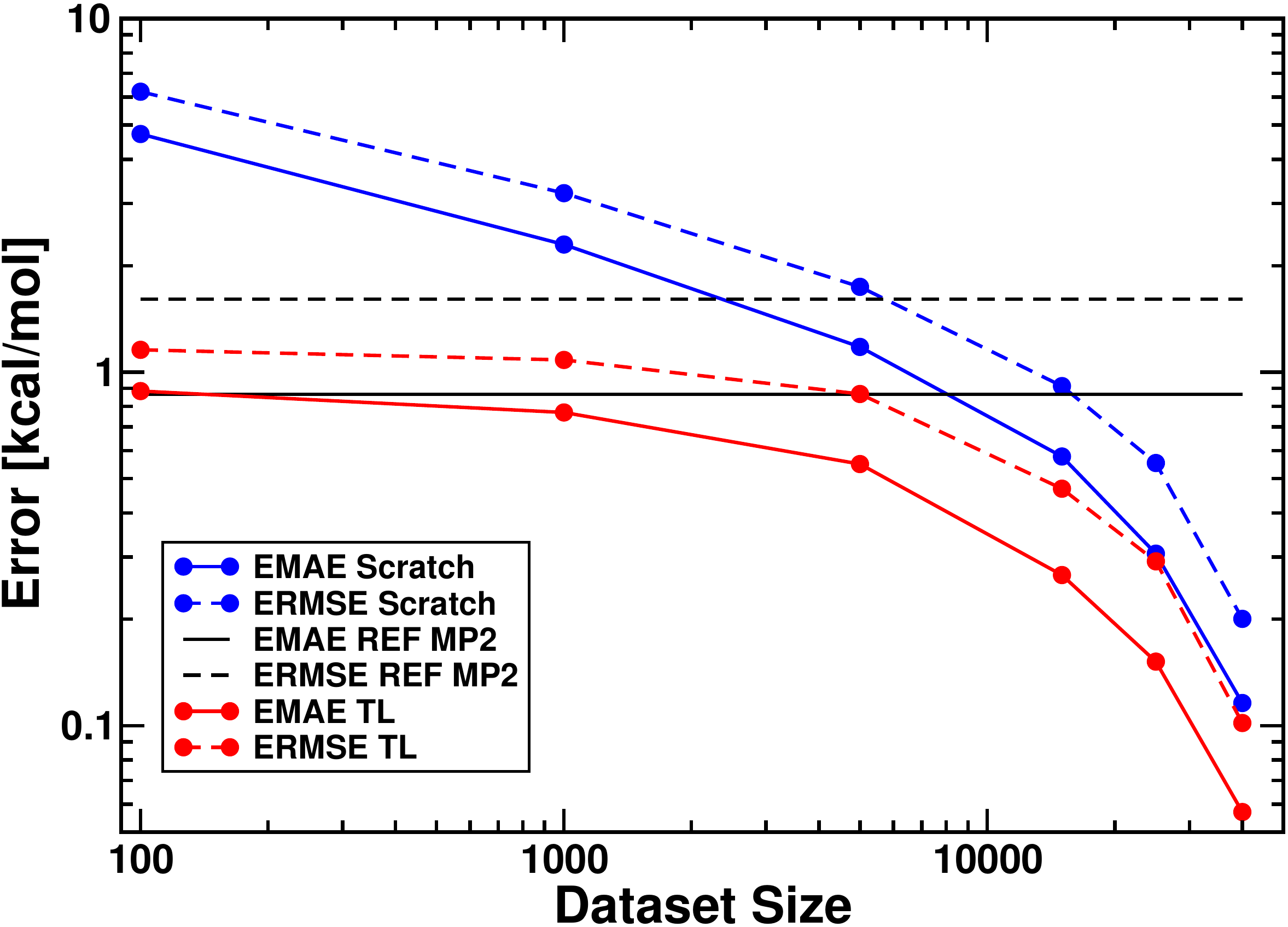}
\caption{Learning curve for the NNs based on PNO-LCCSD(T)-F12 (blue)
  and for the TL models (red). The MAE (solid) and the RMSE (dashed)
  are shown in kcal/mol for all models. All models were evaluated on
  the same test set consisting of 4900 structures. The black lines
  show the performance of the MP2 model trained on 71\,208 structures
  on the same test set.}
\label{fig:transfer_learing_times}
\end{figure}

Transfer learning for the present application requires reference
calculations at a higher level of theory. For this, energies for a
total of 49\,000 geometries for MA, AAA, AcAc and the amons were
calculated at the PNO-LCCSD(T)-F12 level of theory. TL from the lower
(MP2) to the higher (PNO-LCCSD(T)-F12) level of theory was carried out
with different training set sizes (100, 1000, 5000, 15\,000, 25\,000,
40\,000) drawn randomly from the higher-level data set. All models
were then evaluated for the same test set containing 4900 randomly
chosen structures, again including MA, AAA, AcAc and their amons.  The
MAE and the RMSE for predicting the energies of the 4900 test
structures are reported in
Figure~\ref{fig:transfer_learing_times}. The black lines correspond to
the performance of the MP2 NN trying to predict the higher level of
theory energies and act as reference. Moreover, the TL models (red)
are compared to models with the same training set trained from scratch
(blue).\\

\noindent
Evidently, the TL models outperform the models trained from scratch.
Their performance is superior for small data set sizes, because the
training already starts with a good guess for the weights and
biases. For larger sizes, the errors of models trained from scratch
converge towards the TL models, however, even for the largest training
set size the TL models reach higher accuracy. These results suggest
that pre-training on lower level data leads to higher-accuracy models
when TL is applied. With 5000 reference structures the TL 
model performs again with chemical accuracy whereas the model
learned from scratch is inferior by a factor of two. In view of the
final quality of the NN trained on the full MP2 data set, adding 100
structures at the higher level of theory already yields a performance
similar to the reference model. These observations all concern the
global PES.\\

\noindent
It is also of interest to consider particular features of the transfer
learned PES, such as the barrier heights for HT. First, this barrier
was determined at the PNO-LCCSD(T)-F12 level for the three systems. As
no gradients are available for this method, the energies for the
minima and transition state geometries were determined at the
optimized MP2 structures which yields barrier heights for MA, AAA, and
AcAc of 4.04, 3.81, and 3.25 kcal/mol, respectively, which compare
with 4.1 kcal/mol (MA)\cite{bowman2008malonaldehyde} and 3.2 kcal/mol
(AcAc)\cite{howard2015infrared} at the CCSD(T) level of theory. Hence,
compared with the MP2 calculations and the NN learned on MP2 energies
all TS energies are in very good agreement with the highest level of
theory (CCSD(T)).\\

\noindent
Next, the barrier heights $E_{\rm b}$ from the TL models
are considered.  The minimum and TS geometry of each compound are
optimized with the respective model before computing $E_{\rm
  b}$. Barrier heights based on the MP2 optimized geometries are given
in Table~S1 for completeness. TL with different numbers
of randomly chosen structures (between 100 and 40\,000) from the
training set shows that between 15\,000 (for AcAc) and 25\,000 (for MA
and AAA) structures are required to obtain barrier heights within
fractions of one kcal/mol of the PNO-LCCSD(T)-F12 values. This is
primarily due to the fact that the structures for TL were randomly
chosen and that it is not guaranteed that the training set contains
sufficient information about the property (here TS energy) of
interest. If particular features of a PES need to be improved, better
choices will be possible, see further below.\\

\begin{table}[htbp]
\caption{Comparison of the barriers heights for the TL models. Here,
  model m$_n$ corresponds to a single model trained with a training
  set of $n$ randomly chosen structures (including MA, AAA, AcAc and
  amons) and $\left<...\right>$ is the average over four independently
  trained NNs with and without including 100 structures along the MEP of the MA HT. For
  AAA, the isomer having the the H-atom on the unsubstituted side of
  the molecule is the zero of energy and is not listed. $E_{\rm b}$ is
  determined from geometries re-optimized with the corresponding
  model. Moreover, the energy predictions of the NN trained at the MP2
  level of theory, the \textit{ab initio} MP2 and PNO-LCCSD(T)-F12
  barriers and the highest level of theory predictions found in
  literature are
  shown\cite{bowman2008malonaldehyde,howard2015infrared}. All energies
  are in kcal/mol.}
\label{tab:tl_eb}
\begin{tabular}{c|c|c|c|c|c|c}
& \multicolumn{2}{c|}{\textbf{$E_{\rm b}$ MA}} & \multicolumn{2}{c|}{\textbf{$E_{\rm b}$ AAA}} &  \multicolumn{2}{c}{\textbf{$E_{\rm b}$ AcAc}} 
\\\hline
& no MEP           & MEP             & no MEP           & MEP             & no MEP            & MEP              \\ \hline
m$_{100}$                         & 3.18             &                 & 1.99             &                   & 0.92              &                  
\\
m$_{1000}$                        & 5.08             &                 & 3.0              &                   & 2.40              &                  
\\
$\left< {\rm m}_{1000}\right>$    & 4.66             & \textit{3.91}   & 2.74             & \textit{3.35}     & 1.80              & \textit{2.72}    
\\
m$_{5000}$                        & 4.94             &                 & 3.80             &                   & 2.66              &                  
\\
m$_{\rm 15000}$                     & 6.28             &                 & 4.32             &                 & 3.31              &                  
\\
$\left< {\rm m}_{\rm 15000}\right>$ & 6.42             & \textit{4.04}   & 4.20             & \textit{3.97}   & 3.32              & \textit{3.34}    
\\
m$_{25000}$                       & 4.09             &                 & 3.84             &                  & 3.28              &                  \\
m$_{40000}$                       & 4.00             &                 & 3.83             &                  & 3.29              &                  \\ 
\hline
m$_{\rm MP2}$                     & 2.79             &                 & 2.46             &                  & 2.17              &                  \\
MP2                               & 2.74             &                 & 2.47             &                  & 2.18              &                  \\
\textbf{P-LC-F12}          & \textbf{4.04}    & \textbf{}       & \textbf{3.81}    & \textbf{}       & \textbf{3.25}     &                  \\ 
\hline\hline
CCSD(T)                    & 4.1\cite{bowman2008malonaldehyde}              &                 & -         &         & 3.2\cite{howard2015infrared}     
  &

\end{tabular}
\end{table}

\noindent
This was explored in two ways. First, four additional, independent
models were trained, each for a training set size of 1000 and 15\,000
structures and the barrier heights were determined from their
average. For the averaged model based on 1000 training structures
($\left< {\rm m}_{1000}\right>$) an $E_{\rm b} = 4.66$ kcal/mol 
is determined with a
range from 4.31 to 5.08 kcal/mol for the individual models, see
Table~\ref{tab:tl_eb}. For $\left< {\rm m}_{15000}\right>$ this
increases to 6.42 kcal/mol (range from 6.00 to 6.87 kcal/mol). Hence,
the larger training set size did not lead to an improved energy
barrier. For AAA, the barriers range from 2.74 to 4.20 kcal/mol for
$\left< {\rm m}_{1000}\right>$ and $\left< {\rm m}_{15000}\right>$,
respectively, compared with 3.81 kcal/mol from PNO-LCCSD(T)-F12. The
energy barriers of the HT for AcAc were also determined and averaged
values of 1.80 and 3.32 kcal/mol were found for $\left< {\rm
  m}_{1000}\right>$ and $\left< {\rm m}_{15000}\right>$,
respectively. The value of 3.32 kcal/mol compares well with the
PNO-LCCSD(T) value of 3.25 kcal/mol.\\

\noindent
Specifically the failure of $\left< {\rm m}_{15000}\right>$ to
qualitatively describe the H-transfer barrier height for MA is a
motivation to consider enrichment of the training data set with
specific information about this property. Hence, four additional
models each are transfer learned based on a) 100 structures from the
MEP for H-transfer in MA and b) the remaining structures randomly
drawn from (MA, AAA, AcAc, and the amons) as before, for training set
sizes of 1000 and 15000, respectively. Table~\ref{tab:tl_eb} shows
averages of the TL models including the MEP structures for all three
systems. Now $E_{\rm b}$ for MA is 3.91 and 4.04 kcal/mol for $\left<
{\rm m}_{1000}\right>$ and $\left< {\rm m}_{15000}\right>$,
respectively, which compares well with the PNO-LCCSD(T)-F12 reference
value of 4.04 kcal/mol. Even though no dedicated MEP-data was included
for AAA and AcAc, these barrier heights are now also accurately
predicted. Barrier heights of 3.35 (2.72) and 3.97 (3.34) kcal/mol are
found for AAA (AcAc) for the two training set sizes, respectively, and compare with a
PNO-LCCSD(T)-F12 value of 3.81 (3.25) kcal/mol. Hence, including
structures along the MEP for HT in MA not only improves the trained
NNs for MA but for AAA and AcAc, too.\\

\noindent
In summary, TL from a fully-trained NN at the MP2 level of theory by
using additional information of the higher PNO-LCCSD(T)-F12 level of
theory yields improved PESs for the three systems considered at the
level of mean absolute and root mean squared error when using randomly
selected reference data, see
Figure~\ref{fig:transfer_learing_times}. However, when evaluating
particular local features of the TL-NNs, such as the barrier height
for hydrogen transfer or the relative stabilization of two structural
isomers, the results are not particularly accurate and may even lack
convergence towards the correct value with increasing size of the
training set.  Including specific information about the property of
interest was shown to considerably improve this. This is reminiscent
of the ``morphing potential'' approach\cite{MM.morphing:1999} which
aims at reshaping a fully-dimensional PES calculated at a lower level
of theory by means of a generalized coordinate transformation and
reference data at a higher level of theory.\\

\section{Conclusion and Outlook}
The PESs of MA, AAA and AcAc are modeled successfully with a single
NN, which predicts the energies of a test set containing 9208
structures, including MA, AAA, AcAc and substructures, with a MAE of
$\approx 0.020$~kcal/mol and a RMSE of $\approx 0.21$~kcal/mol. The NN
based on MP2/aug-cc-pVTZ calculations is able to predict the energy
barriers for the three HT reactions with deviations $\leq 0.05$
kcal/mol in comparison to the MP2/aug-cc-pVTZ values. The model can be
used to run simulations, calculate IR spectra or reaction rates.
Both, the IR spectra and the rates were found to agree with
literature. The generalizability of the NN was examined with a
prospect to simplify future research in terms of modeling. It was
shown that the NN is able to generalize fairly well to structures with
an extra methyl group, although not all properties will be predicted
perfectly. Depending on what the aim of the theoretical investigation
is, this approach could be used to circumvent considerable amounts of
computing time.\\

\noindent
Finally, the method of TL was explored and the MP2 NN was transfer
learned to the PNO-LCCSD(T)-F12 level of theory. First, models with
different training set sizes are used to predict the energies of the
test set containing 4900 structures. The TL models outperform the
models trained from scratch in terms of learning curve and accuracy.
Second, the energy barriers of the HT reactions are examined and it is
found that between 15\,000 (for AcAc) and 25\,000 (for MA and AAA)
random structures are required to obtain barrier heights within
fractions of one kcal/mol of the PNO-LCCSD(T)-F12 values. Including
structures along the MEP for HT in MA considerably improves the
barrier heights found in the TL models for MA, AAA, and AcAc, with a
maximal difference of $0.26$~kcal/mol between the NN and the reference
calculations.\\

\noindent
In summary, a comprehensive NN model for MD studies of the
spectroscopy and HT dynamics for MA, AAA, and AcAc was trained at the
MP2 level of theory. The findings further suggest that TL will provide
an efficient route forward for high-level, fully dimensional and
reactive models to investigate the dynamics of chemical systems. Also,
extending chemical space is expected to be possible but more
systematic studies on this aspect are required.\\

\section*{Acknowledgments}
Financial support by the Swiss National Science Foundation through
grants 200021-117810, the NCCR MUST, and the University of Basel is
acknowledged. OTU acknowledges funding from the Swiss National Science
Foundation (Grant No. P2BSP2\_188147).

\bibliographystyle{unsrt}

\bibliography{references}

\end{document}

\section{Reference Structure Set}
\begin{figure}[h!]
\centering \includegraphics[width=0.8\textwidth]{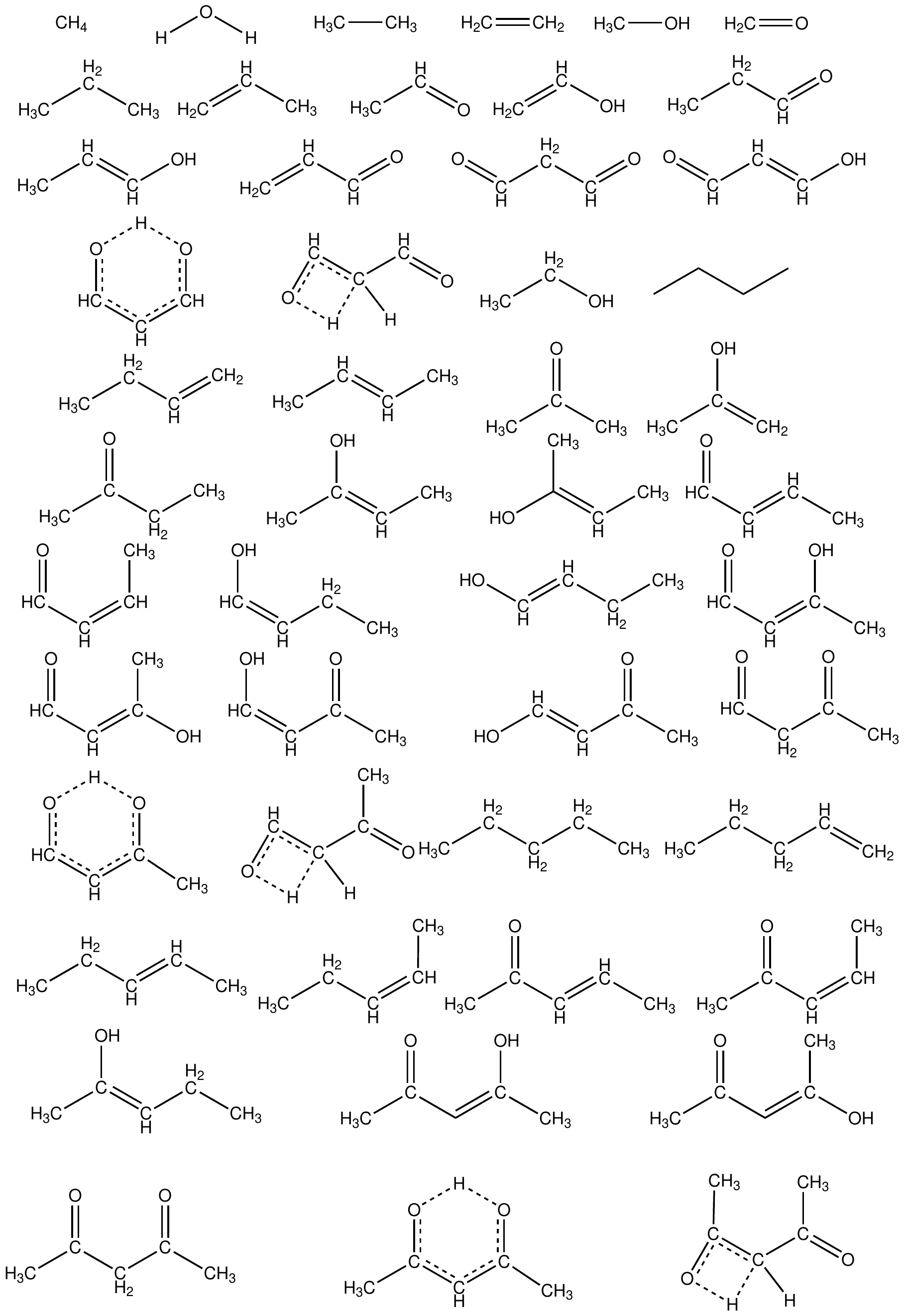}
\caption{The complete reference structure set used in the present work.}
\label{sifig:amons}
\end{figure}

\section{Analysis of the High-Dimensional Potential Energy Surface}
\begin{figure}[htbp]
\centering
\includegraphics[width=0.9\textwidth]{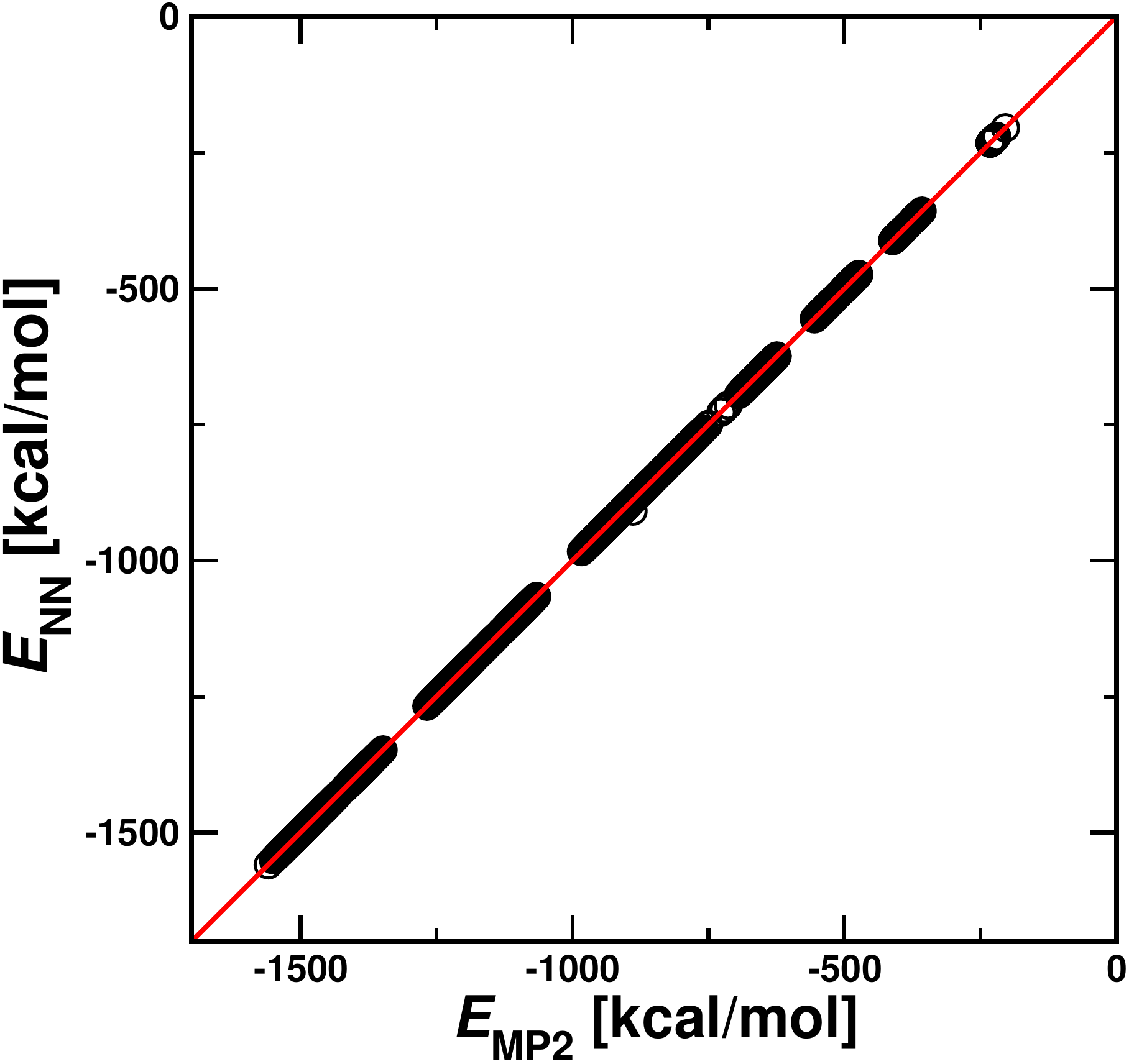}
\caption{Comparison of the MP2/aug-cc-pVTZ reference and predicted NN
  energies for the test set containing MA, AAA, AcAc and their amons.
  The data from the test set (9208 structures), which were not used in
  training the NN are predicted by the NN and compared with the
  reference energies from MP2/aug-cc-pVTZ calculations. The mean
  absolute error is 0.021 kcal/mol.}
\label{sifig:corr_ma_final_a}
\end{figure}
\clearpage
\section{Proton Transfer Rates}\label{sec:si_proton_transfer_rates}
\begin{figure}[htbp]
\centering
\includegraphics[width=0.9\textwidth]{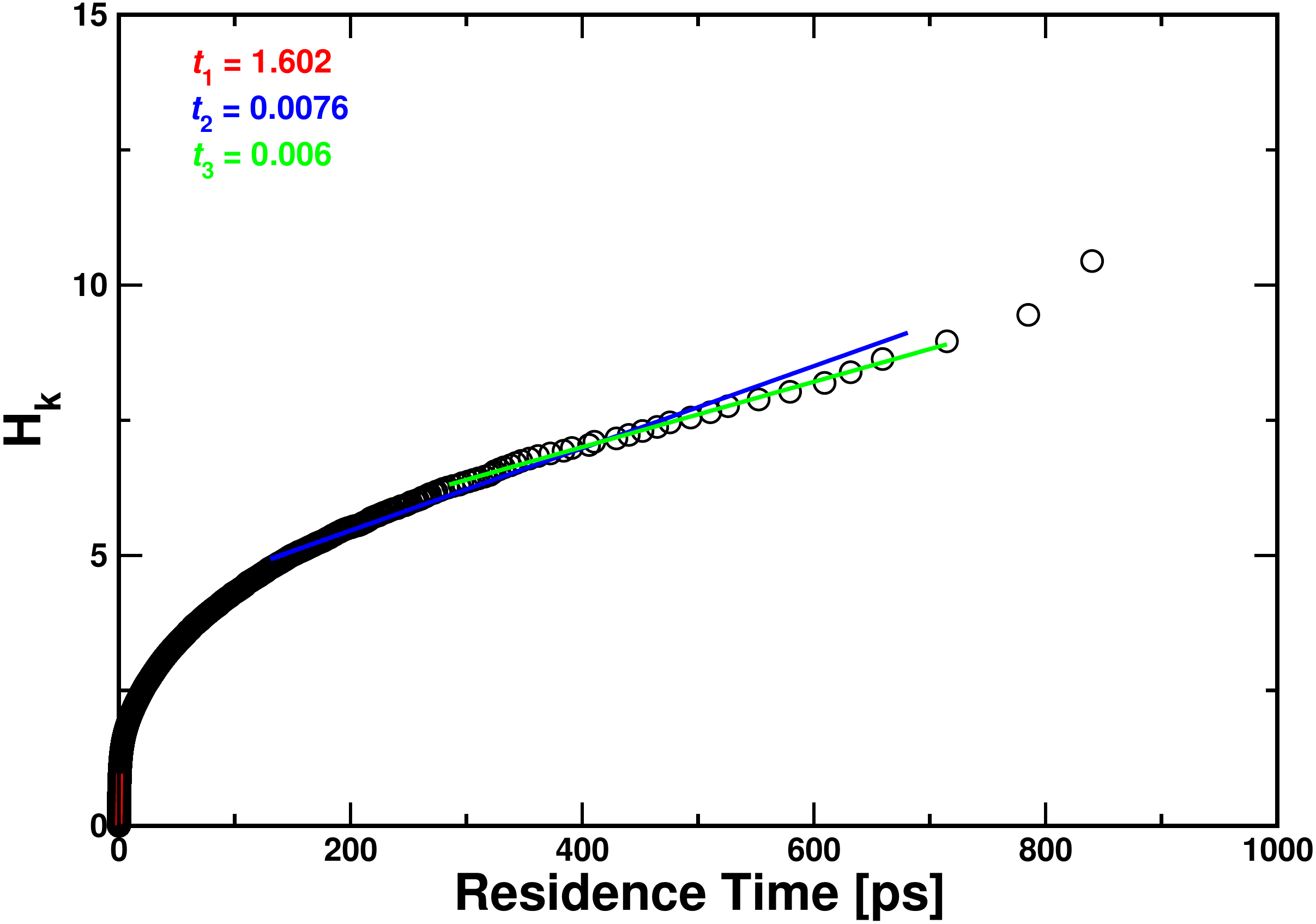}
\caption{The hazards for the MA trajectories for $r_c= 1.1$
  \r{A} (only every fifth data point
  is shown). Depending on the range of the long-time data included in the
  fit, the long time scale $t_2$ can vary between 0.0060 and 0.0076
  transitions/ps.}
\label{sifig:hazard_rc_1.1}
\end{figure}

\begin{figure}[htbp]
\centering
\includegraphics[width=0.9\textwidth]{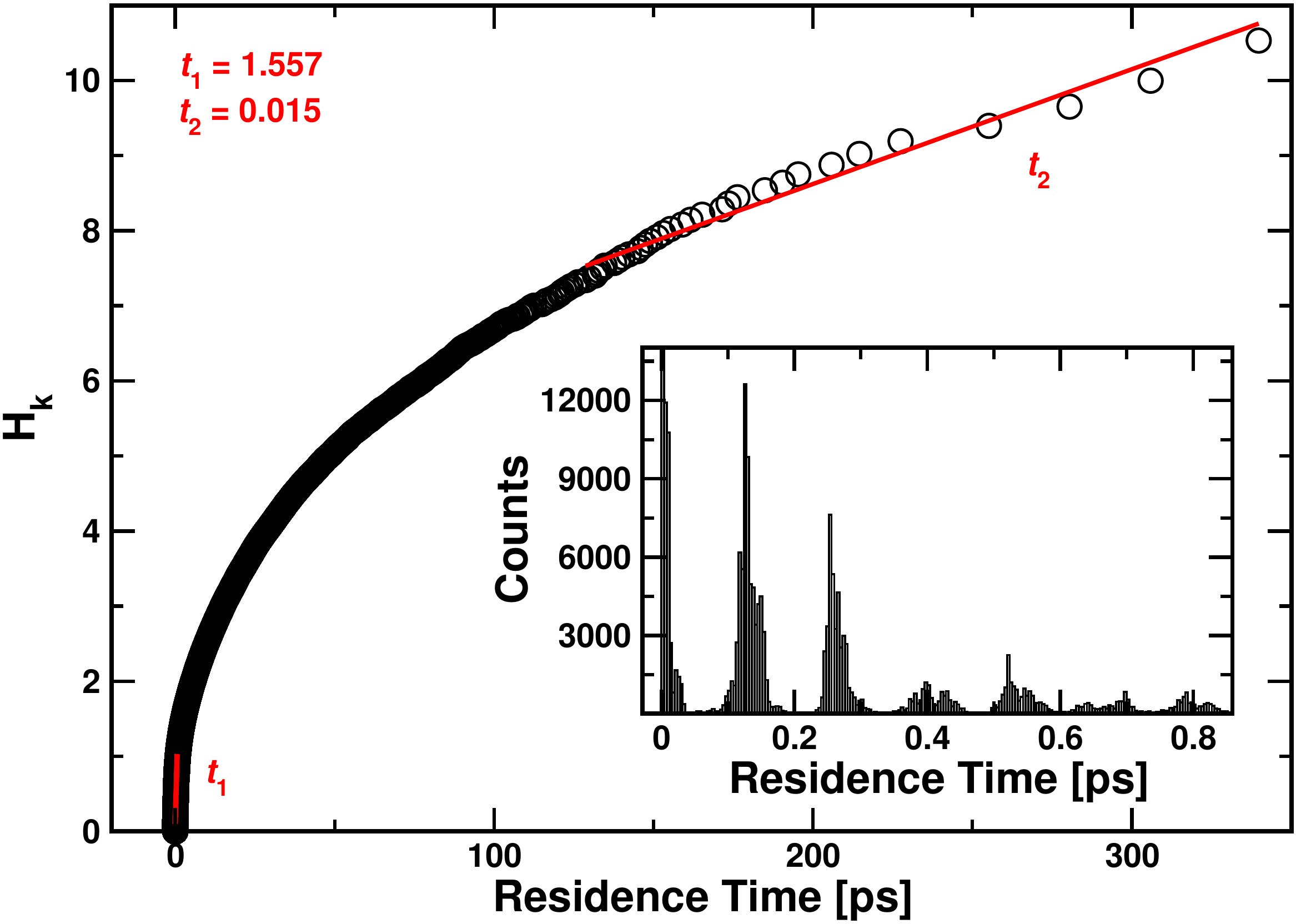}
\caption{Hazard plot and residence times (inset) for a total of 1
  $\mu$s simulation for AAA structure at 300 K (only every fifth data point
  is shown). About 60 \% of the
  events concern residence times $\tau < 0.6$ ps and the longest
  residence time is $\approx 960$ ps. A linear regression for the fast
  and slow process yield rates of $t_1 = 1.557$ crossings/ps and $t_2
  = 0.015$ crossings/ps. The inset illustrates the histogram of the
  residence times with a period of $\approx 0.135$ ps, corresponding
  to a frequency of 247 cm$^{-1}$.}
\label{sifig:met_ma_hazard}
\end{figure}

\begin{figure}[htbp]
\centering
\includegraphics[width=0.9\textwidth]{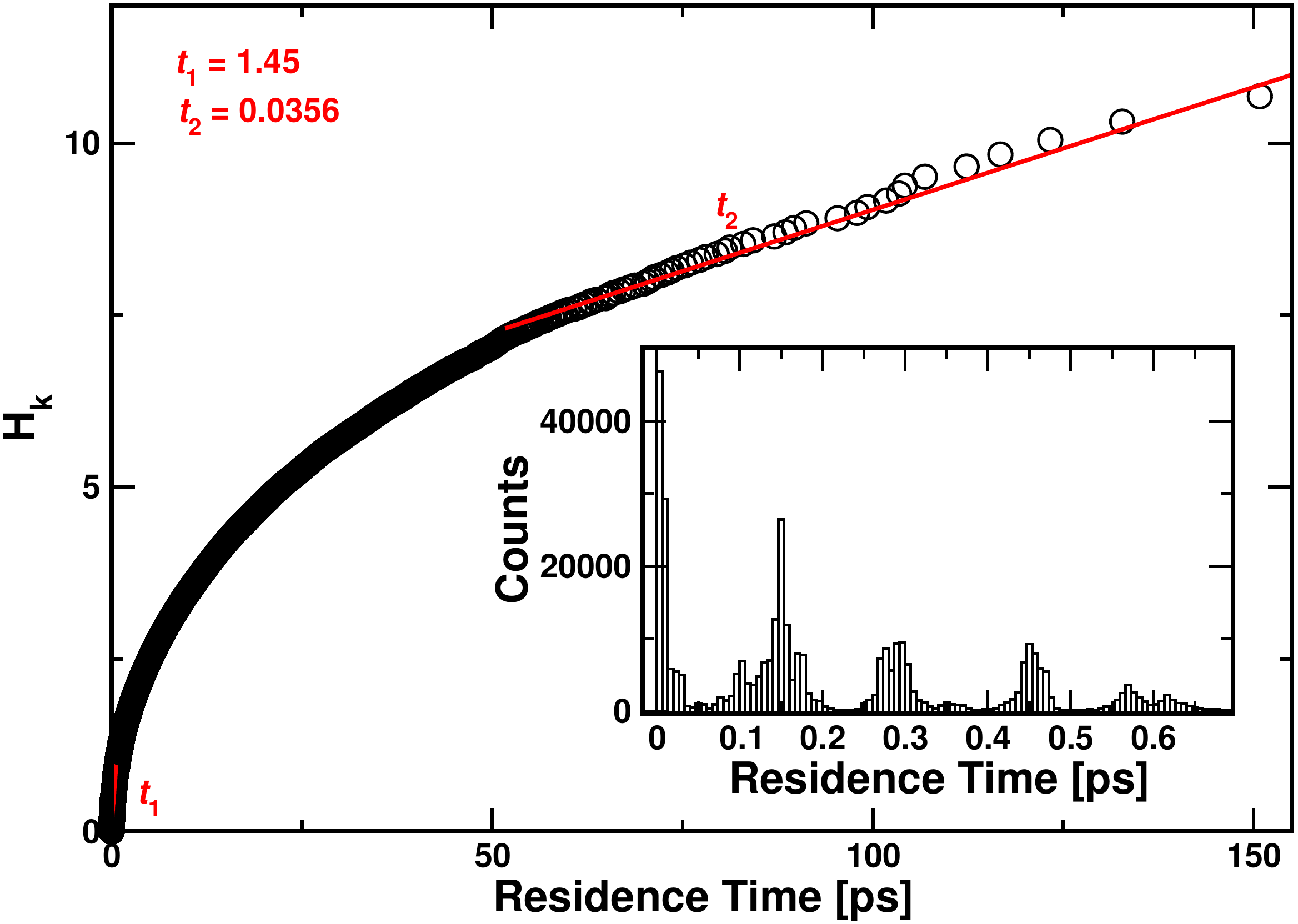}
\caption{Hazard plot and residence times (inset) for a total of 1
  $\mu$s simulation for AcAc structure at 300 K (only every fifth data point
  is shown). About 60 \% of the
  events concern residence times $\tau < 0.6$ ps and the longest
  residence time is $\approx 260$ ps. A linear regression for the fast
  and slow process yield rates of $t_1 = 1.45$ crossings/ps and $t_2 =
  0.036$ crossings/ps. The inset illustrates the histogram of the
  residence times with a period of $\approx 0.15$ ps, corresponding to
  a frequency of 222 cm$^{-1}$.}
\label{sifig:acac_hazard}
\end{figure}
\clearpage
\section{Generalizability of the Neural Network}
\begin{figure}[htbp]
\centering
\includegraphics[width=0.8\textwidth]{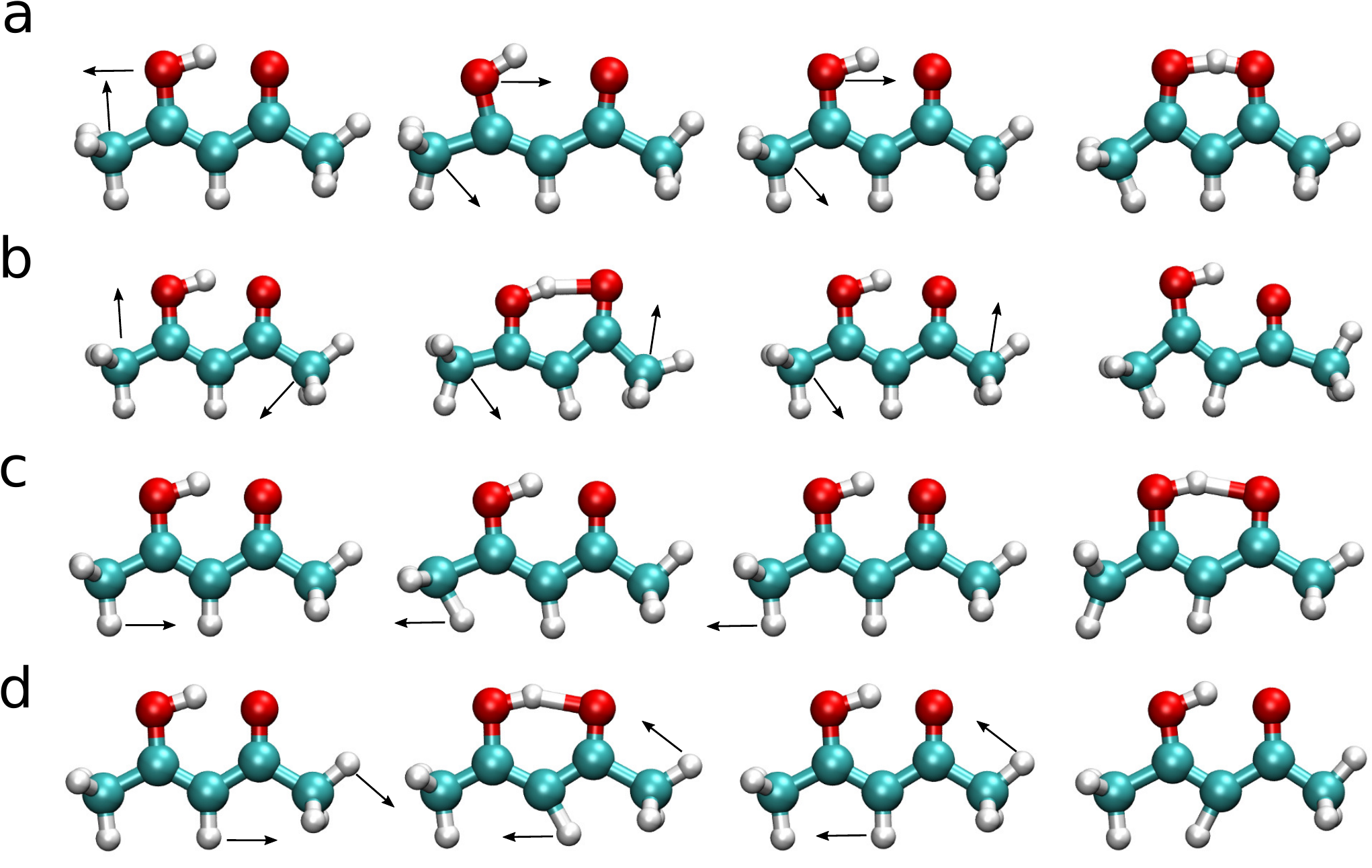}
\caption{Illustration of the modes with the largest deviation between
  NN and NN* for AcAc. The black arrows show the displacement
  direction of the atoms. The modes shown in a, b, c, d correspond to
  the modes at 367 (381), 397 (407), 1032 and 1187 (1176) cm$^{-1}$
  for the NN* model.}
\label{sifig:modes}
\end{figure}
\clearpage
\section{Transfer Learning to a Higher Level of Theory}
\begin{table}[htbp]
\caption{Comparison of the barriers heights for the TL models. Here,
  model m$_n$ corresponds to a single model trained with a training
  set of $n$ randomly chosen structures (including MA, AAA, AcAc and
  amons). In contrast to Tab.~3 in the main manuscript
  the energy barriers are reported for the MP2 optimized
  geometries. Moreover, the \textit{ab initio} MP2 and
  PNO-LCCSD(T)-F12 barriers and the highest level of theory
  predictions found in literature are
  shown\cite{bowman2008malonaldehyde,howard2015infrared}. All energies
  are in kcal/mol.}
\label{sitab:tl_eb}
\begin{tabular}{c|c|c|c}
& \textbf{$E_{\rm b}$ MA} & \textbf{$E_{\rm b}$ AAA} &  \textbf{$E_{\rm b}$ AcAc} \\\hline
m$_{100}$		&2.75	&	1.80	&	0.87	\\
m$_{1000}$	&5.17	&	3.26	&	2.54	\\
m$_{5000}$	&4.62	&	3.55	&	2.66	\\
m$_{15000}$	&6.14	&	4.17	&	3.22	\\
m$_{25000}$	&4.05	&	3.76	&	3.23	\\
m$_{40000}$	&3.97	&	3.79	&	3.22	\\\hline

MP2                               & 2.74                             & 2.47                              & 2.18    \\
\textbf{P-LC-F12}          & \textbf{4.04}    & \textbf{3.81}    & \textbf{3.25} \\ 
\hline\hline
CCSD(T)                    & 4.1\cite{bowman2008malonaldehyde}              & -    & 3.2\cite{howard2015infrared}

\end{tabular}
\end{table}

\clearpage
\bibliographystyle{unsrt}
\bibliography{references}